\documentclass[sigconf, nonacm]{acmart}
\usepackage{amsmath}
\usepackage{graphicx}
\usepackage{textcomp}
\usepackage{xcolor}
\usepackage{balance}
\usepackage{hyperref}
\usepackage{subfigure}

\usepackage{listings}
\usepackage{xspace}
\usepackage{color}
\usepackage{booktabs}

\newcommand\doi{XX.XX/XXX.XX}
\newcommand\availabilityurl{URL_TO_YOUR_ARTIFACTS}

\hypersetup{hidelinks}
\newcommand{\wrt}{\emph{w.r.t.}\xspace}
\newcommand{\ie}{\emph{i.e.,}\xspace}
\newcommand{\eg}{\emph{e.g.,}\xspace}
\newcommand{\resp}{\emph{resp.,}\xspace}

\newcommand{\eat}[1]{}
\newcommand{\btitle}[1]{\vspace{1ex}\noindent \textbf{#1}}
\newtheorem{definition}{Definition}

\usepackage{algorithm}  
\usepackage{algorithmicx}  
\usepackage{algpseudocode}
\begin{document}

\title{One Size Cannot Fit All: a Self-Adaptive Dispatcher for Skewed Hash Join in Shared-nothing RDBMSs
}

\author{Jinxin Yang}
\author{Hui Li}
\author{Yiming Si}
\affiliation{%
  \institution{Xidian University}
  \city{Xi'an}
  \country{China}
}
\email{yangjx@stu.xidian.edu.cn}
\email{hli@xidian.edu.cn}
\email{Si\_ym@hotmail.com}

\author{Hui Zhang}
\author{Kankan Zhao}
\author{Kewei Wei}
\affiliation{%
  \institution{Inspur Software Group Co. Ltd.}
  \city{Beijing}
  \country{China}
}
\email{zhanghui@inspur.com}
\email{zhaokankan@inspur.com}
\email{weikewei@inspur.com}

\author{Wenlong Song}
\author{Yingfan Liu}
\author{Jiangtao Cui}
\affiliation{%
  \institution{Xidian University}
  \city{Xi'an}
  \country{China}
}
\email{wlsong\_xdu@foxmail.com}
\email{{liuyingfan,cuijt}@xidian.edu.cn}

\begin{abstract}
Shared-nothing architecture has been widely adopted in various commercial distributed \textsc{rdbms}s. Thanks to the architecture, query can be processed in parallel and accelerated by scaling up the cluster horizontally on demand. In spite of that, load balancing has been a challenging issue in all distributed \textsc{rdbms}s, including shared-nothing ones, which suffers much from skewed data distribution. 
In this work, we focus on one of the representative operator, namely \textsf{Hash Join}, and investigate how skewness among the nodes of a cluster will affect the load balance and eventual efficiency of an arbitrary query in shared-nothing \textsc{rdbms}s. We found that existing Distributed \textsf{Hash Join} (\textsf{Dist-HJ}) solutions may not provide satisfactory performance when a value is skewed in both the probe and build tables. 

To address that, we propose a novel \textsf{Dist-HJ} solution, namely Partition and Replication (PnR). Although PnR provide the best efficiency in some skewness scenario, our exhaustive experiments over a group of shared-nothing \textsc{rdbms}s show that there is not a single \textsf{Dist-HJ} solution that wins in all (data skew) scenarios. To this end, we further propose a self-adaptive \textsf{Dist-HJ} solution with a built-in sub-operator cost model that dynamically select the best \textsf{Dist-HJ} implementation strategy at runtime according to the data skew of the target query. We implement the solution in our commercial shared-nothing \textsc{rdbms}, namely KaiwuDB (former name ZNBase) and empirical study justifies that the self-adaptive model achieves the best performance comparing to a series of solution adopted in many existing \textsc{rdbms}s.
\end{abstract}

\maketitle

\pagestyle{plain}
\begingroup\small\noindent\raggedright\textbf{Reference Format:}\\
\authors. \shorttitle\\
\href{https://doi.org/\doi}{doi:\doi}
\endgroup
\begingroup
\renewcommand\thefootnote{}\footnote{\noindent
This work is licensed under the Creative Commons BY-NC-ND 4.0 International License. Visit \url{https://creativecommons.org/licenses/by-nc-nd/4.0/} to view a copy of this license. For any use beyond those covered by this license, obtain permission by emailing . Copyright is held by the owner/author(s). Publication rights licensed to the . \\
\href{https://doi.org/\doi}{doi:\doi} \\
}\addtocounter{footnote}{-1}\endgroup

\ifdefempty{\availabilityurl}{}{
\vspace{.3cm}
\begingroup\small\noindent\raggedright\textbf{Artifact Availability:}\\
The source code, data, and/or other artifacts have been made available at \url{https://github.com/lihuixidian/SkewHJs}.
\endgroup
}

\section{Introduction}
\begin{figure}
  \centering
  \includegraphics[width=\linewidth,height=6cm]{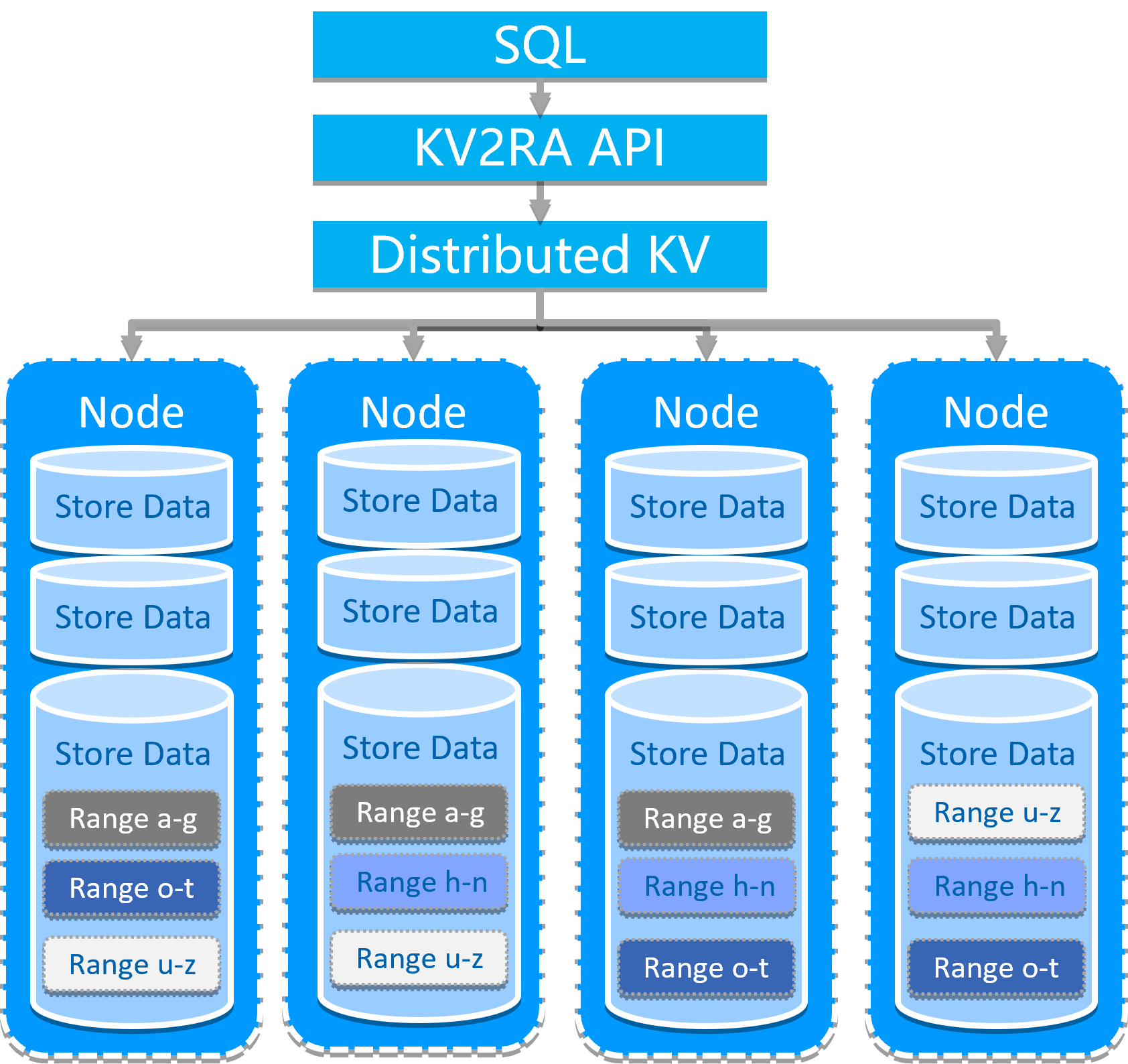}
  \vspace{-2ex}\caption{Architecture of KaiwuDB (formerly ZNBase)}
  \label{fig:znbase}\vspace{-4ex}
\end{figure}

Due to the native support of parallel query processing and ease of scaling, shared-nothing architecture has been widely adopted in various commercial distributed \textsc{rdbms}s~\cite{2020CockroachDB,2020Ti}. These \textsc{rdbms}s employ the sharding technique and partition data horizontally across a database cluster. Parallel query processing can be easily accelerated simply by adding servers. Following the shared-nothing~\cite{stonebraker1986case} architecture, Inspur Software Group developed a new commercial \textsc{rdbms}s, namely KaiwuDB (former name ZNBase\footnote{https://www.znbase.com/website/}), the system architecture of which is shown in \autoref{fig:znbase}. Each single node of KaiwuDB has independent computational, memory and storage resources, where the storage engine is built upon a popular KV-store, namely RocksDB~\cite{DBLP:conf/cidr/DongCGBSS17}. 

Tables are stored within a monolithic map of key-value pairs. Keys of a single table ranges contiguously in the key space, according to which a table is partitioned into several contiguous slices, namely ``ranges''. A range is replicated into 3 (by default) replicas, which are accordingly stored on 3 different nodes. Among all the replicas, there is one called ``leaseholder'', who is in charge of coordinating read and write requests for the range. Ideally, read requests are always sent to leaseholders, the tuples of which are always kept up-to-date. Notably, the leaseholders are non-static, and is elected and managed according to Raft protocol~\cite{2014In}. Similar settings are popularly adopted in most shared-nothing \textsc{rdbms}s~\cite{2020CockroachDB,2020Ti}.
 
\subsection{\textsf{Dist-HJ} in KaiwuDB}
Given an SQL query, an \textsc{rdbms} compiles it and correspondingly generates a  query execution plan (\textsc{qep}), with the help of a query optimizer. A \textsc{qep} consists of a group of physical operators, which are organized in a tree structure, showing how the given query will be processed and answered step-by-step (\ie operator-by-operator). Traditional \textsc{rdbms} contains tens of predefined operators, among which \textsf{Hash Join} is the one that are popularly used and studied~\cite{shafieinejad2022equi,mageirakos2022efficient,tziavelis2022toward,zhang2021fast,zhang2022learning,dossinger2021optimizing,muller2021memory}. Generally, a \textsf{Hash Join} operator consists of a pair of sub-steps including hash redistribution and join; specifically, within a shared-nothing distributed (\eg KaiwuDB) \textsc{rdbms}, \textsf{Hash Join} has to be executed in a distributed way. Similar to~\cite{2020CockroachDB,2020Ti}, given an SQL statement, the query plan generation within KaiwuDB can be generally divided into two phases. In the first phase, called \textit{logical planning}, the corresponding Abstract Syntax Tree (\textsc{ast}) parsed from the given SQL is fed into the optimizer to generate the optimized logical plan. A logical plan is structured into a tree, where a node represents a logical operator. In the second phase, called \textit{physical planning}, the system transforms the optimized logical plan into a physical plan, which specifies how executions are \textit{distributed} and performed in parallel over the cluster. For instance, a logical \textsf{SCAN} operation (generated in the first phase) is carried out by a series of \textsf{Tablereader}s (physically), one for each node. \textsf{Tablereader} merges the access to ranges that are on the same node. Similarly, subsequent logical operators, \eg \textsf{JOIN}, will also be implemented physically and distributed at each node, \eg \textsf{Hash Join}, in this phase. 

As the locality of ranges are interleaved across nodes (\ie spanned over some nodes without strictly following an order), yet the distribution of the ranges are not static (\ie may vary when the tuples are changed), physical alignment of the \textsf{Tablereader}s is dynamically decided at runtime. Accordingly, as a subsequent operator, the specific workload of distributed \textsf{Hash Join} (\textsf{Dist-HJ} for short) highly depends on what the \textsf{Tablereader}s at the earlier step output. 
 \begin{figure}
  \centering
  \includegraphics[width=\linewidth,height=3.5cm]{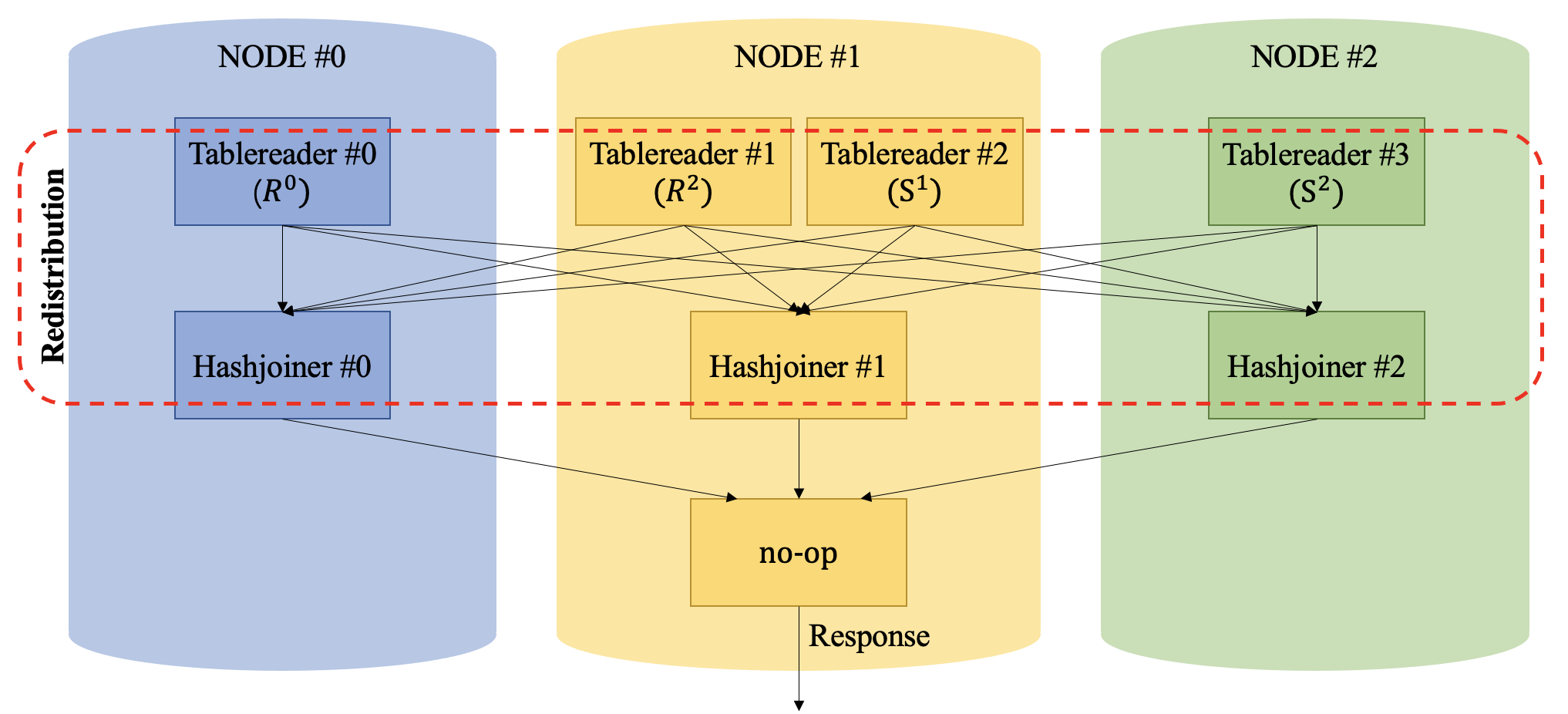}
  \vspace{-3ex}\caption{A potential execution plan of \textsf{Dist-HJ} in KaiwuDB}
  \label{fig:executionplan}\vspace{-2ex}
\end{figure}
\vspace{-2ex}\subsection{Data Skew in \textsf{Dist-HJ}}\label{subsec12} 
\autoref{fig:executionplan} reveals a potential execution plan of \textsf{Dist-HJ} in KaiwuDB running on a cluster with $k\ge 3$ nodes. Without loss of generality, suppose that an SQL task as follows is carried out over 3 nodes in the cluster and thus the \textsf{Dist-HJ} will be executed over the 3 nodes. 
\begin{lstlisting}[language=SQL,keywordstyle=\color{blue},xleftmargin=.05\textwidth,xrightmargin=.05\textwidth
]
SELECT * FROM R,S WHERE R.a=S.b;  
\end{lstlisting}    
According to the distributed execution plan, partitions of $R$ are accessed on node 0 and 1, and partitions of $S$ are accessed on node 1 and 2, respectively. By default, primitive hash redistribution is employed further to distribute the join task over the nodes such that it is able to run in parallel~\cite{DBLP:conf/icde/KitsuregawaTN92}. 

Assume that there exists a skewed value $x$ in $R.a$, and $hash(x)\bmod{3}=1$. Formally, given a table $R$, a value $x$ in column $a$ is \textbf{skewed} if it appears in more than $p|R|$ (\eg 20\%) entries in an arbitrary column, where $p$ is referred to as the \textit{skewness threshold}. Consequently, node 1 will be overloaded, as the tuples with $R.a=x$ are all assigned to it. 
In the literature, there are two representative solutions to address skewed \textsf{Hash Join}, namely Partial Redistribution and Partial Duplication (PRPD)~\cite{2008Handling} and Flow-Join~\cite{2016Flow}. Under PRPD, tuples with value $x$ from $R$ are kept locally on node 0 and node 1, and tuples with $x$ from $S$ on node 1 and node 2 are broadcasted to all the three nodes. The intuition of PRPD is to balance the workloads of all three nodes. However, keeping the skewed tuples of $R$ locally will inevitably lead to under-utilization of node 2, as there is no skewed tuples of $R$ initially aligned to node 2. On the other hand, for Flow-Join, if $x$ is not skewed in $S$, it behaves the same as PRPD. Otherwise, Symmetric Fragment Replicate (SFR)~\cite{stamos1993symmetric} is employed to balance the workloads among nodes. However, as we shall see later in this work, SFR also fails to balance the workloads in many practical scenarios. 

In particular, skewness for \textsf{Dist-HJ} in KaiwuDB can be caused by either nonuniform data distribution or biased physical planning. Intuitively, for both cases, suppose that the skewed tuples from $R$ can be assigned in round-robin fashion to all nodes and the correlated tuples from $S$ can be broadcast, then all nodes may receive approximately the same number of skewed tuples from both tables. Therefore, it balances the workload among nodes. Driven by the above insight, we make a new attempt in KaiwuDB by proposing a new \textsf{Dist-HJ} solution, namely Partition and Redistribution (PnR).

Notably, compared with PRPD and Flow-Join, as round-robin is adopted in PnR, many skewed tuples from $R$ are now distributed over the network, which may introduce extra communication cost. In practice, we discovered that none of these solution wins all in KaiwuDB (our empirical and theoretical study in this work show that the phenomenon exist not only in KaiwuDB but also in other shared-nothing \textsc{rdbms}s). It directly motivates us to develop a self-adaptive solution further in practical system. PnR and Self-Adaptive solution are unfolded in detail in the rest of this paper.

The rest of this paper is organized as follows. In \autoref{sec:rel}, we review and categorize a list of representative solutions towards \textsf{Dist-HJ}, showing their pros and cons in face of different skew scenarios. Afterwards, we theoretically discuss and compare the strategies adopted in GraHJ (the default solution in KaiwuDB and a series of other \textsc{rdbms}s~\cite{2020CockroachDB,lyu2021greenplum}) and PRPD~\cite{2008Handling} (one of the state-of-the-art solution for skewness in \textsf{Dist-HJ}) towards skewed values \textsf{Dist-HJ} and uncover the limitations of both. Accordingly, we propose a novel strategy as an alternative, namely Partition and Redistribution (PnR) in \autoref{sec3}. We implement all three strategies (viewed as different sub-operators of \textsf{Dist-HJ}) in KaiwuDB, and find (via both theoretical study and empirical test) that none of them can win in all skew scenarios. To address that issue, in \autoref{sec:4} we propose a self-adaptive sub-operator dispatcher that is native embedded into the query optimizer and dynamically finds the best strategy by identifying the skew scenario at runtime. We implement the proposed self-adaptive dispatcher in KaiwuDB and test the performance in~\autoref{sec:5}. Finally, \autoref{sec:6} concludes the contributions in this paper.

\section{Related studies and systems}\label{sec:rel}
Data skew has been extensively studied in parallel query processing within distributed \textsc{rdbms}s. Data skew is generally classified into four types in \cite{dewitt1992practical}, which are Tuple Placement Skew (TPS), Selectivity Skew (SS), Redistribution Skew (RS) and Join Product Skew (JPS). TPS refers to the scenario that the initial distribution of tuples among $N$ replicas is unbalanced, which is avoidable by carefully choosing partition columns and placement of replicas. SS refers to the case caused by the different selectivities of filtering predicates, which is rarely studied because the predicates vary in queries. RS refers to the imbalance in the number of the received data for different nodes after redistribution. JPS refers to the skew of join results of each nodes. Most of the studies focus on RS and JPS. In this work, we focus on the RS type, as redistribution is a key step that is the most sensitive to skewness in \textsf{Dist-HJ}. In the followings, we shall investigate several representative solutions in this regard. 

Prior studies \wrt RS can be classified into three categories. The first category of algorithms consider the skew problem as a conventional task scheduling job, in which two join tables are scanned and materialized for a specific analysis stage and both tables are partitioned into smaller units accordingly.  \cite{dewitt1992practical} introduces range partitioning to replace hash partitioning in order to balance the volume of tuples each node receives. Range partitioning is based on a $\emph{split vector}$ computed in a separate sampling stage. Hash-based method is used in \cite{2001Skew,1991Handling,kitsuregawa1990bucket}, the build and probe table are partitioned into several buckets using hash partitioning. The partitioned units are distributed via task scheduling models like LPT~\cite{1969Bounds} and MULTIFIT~\cite{1978An}. Although the task scheduling algorithms may alleviate the imbalance during redistribution step, it introduce significant overhead to the join step, even when the tables are non-skewed. The second category of algorithms dynamically handle the skew problem \cite{1995Dynamic,1993Using,1995Handling}. Overloaded nodes migrate exceeding workloads to other idle nodes. In the shared-nothing system like KaiwuDB, there is no task stealing between nodes due to the following challenges: (a) The elegant shared-nothing architecture, where each node has independent storage and computation\cite{1999Scheduling}, requires restructuring to support task stealing. (b) Task stealing incurs additional system overhead, as nodes need extra time for communication and coordination. This can increase latency and decrease the overall system throughput, even when using remote direct memory access (RDMA), which still requires a series of one-sided RDMA communications that cause high stealing latency\cite{2021Optimizing}. (c) Task stealing may also result in a significant large load on the node where the task is migrated to~\cite{2021DS}. The third category of algorithms is the selective-broadcast solution. The selective-broadcast algorithm is firstly proposed in \cite{dewitt1992practical}, which is known as $\emph{subset-replication}$. The key idea of selective-broadcast algorithm is to keep skewed tuples in one table locally, and meanwhile replicate and broadcast the correlated tuples in the other table. The first two category of algorithms solve the skew problem at the macro-level, without combining the characteristics of Join algorithm. In comparison, the processing method for skew value in the selective-broadcast algorithms is more targeted and easier to adapt to the architecture of KaiwuDB. We review two representative selective-broadcast algorithms, namely PRPD~\cite{2008Handling} and Flow-Join~\cite{2016Flow}, in detail in the followings. 
\vspace{-1ex}\subsection{Partial Redistribution Partial Duplication}
PRPD assumes that $R$ and $S$ are partitioned using a carefully chosen column, such that each node holds almost the same number of both skewed and non-skewed tuples in the same table. PRPD takes advantage of the statistics and assumes that the skewed value sets of each table is known in advance. The key idea of PRPD is to handle the skewed tuples and non-skewed ones separately. In particular, non-skewed tuples are normally hash redistributed, while skewed tuples in one table are kept locally (at the node where it is generated) and the corresponding tuples in the other table are broadcast. Accordingly, each table is divided into three disjoint sets before the join, \ie tuples to be hash redistributed, kept locally and broadcast, respectively. Joins are performed on the these sets after redistributed/broadcast to the particular nodes, the union of whose results constructs the output of \textsf{Dist-HJ}.  
PRPD has also proposed a variation (referred to as PRPD-u\cite{2008Handling}) towards the unevenly distributed case, where most of skewed data are on one or a few nodes in the cluster. To illustrate that, assume one table is evenly partitioned and most of its skewed tuples are located on one ``hot node''. Following PRPD-u, the skewed data on each node are \textit{randomly redistributed} to other nodes instead of being kept locally. For the other table, the redistribution plan remains unchanged. This \textit{random redistribution} is claimed to be able to approximately balance the workloads of all nodes.

\vspace{-1ex}\subsection{Flow-Join}\label{ssec22}
Different from PRPD, Flow-Join proposes to detect the skewed values at run time instead of using statistics in advance. Firstly, Flow-Join dynamically maintains a small histogram to identify the skewed data. According to the empirical results of Flow-Join, the time cost of the detection work does not exceed $2\%$ of the overall execution time.  Secondly, after obtaining the (estimated) skewed values for both input ($R$ and $S$), Flow-Join selects to follow the same strategy as PRPD: the skewed values of one input are kept local and the corresponding values in the other table are broadcast. Aside from the above similarity, the way to handle the skewed tuples, \ie broadcast, after the runtime detection is different in Flow-Join. Flow-Join redistributes skewed values via the Symmetric Fragment Replicate (SFR) redistribution scheme. In fact, if we turn off the runtime skew detector by relying on collected statistics and disable SFR, both PRPD and Flow-Join behave the same. Therefore, PRPD and Flow-Join are uniformly referred to as PRPD in the rest of this paper.

Since statistics towards skewed values are easy to obtain (via collected frequency and histograms), the selective-broadcast algorithm is a promising solution to address skewness in \textsf{Dist-HJ} for KaiwuDB. However, as discussed in Section~\ref{subsec12}, as physical plan is dynamically generated by taking into account a series of runtime load over the cluster, the volume and distribution of data accessed by \textsf{Tablereader}s is not constant, even for the same query. Consequently, TPS may exists in \textsf{Dist-HJ} in KaiwuDB and the skewness of TPS may fluctuate significantly. It does not satisfy the assumption the majority of existing solutions have made that the build and probe tables (in \textsf{Hash Join}) are evenly accessed at each node. In this paper, we present a novel algorithm called PnR. PnR alleviates the JPS caused by correlated tuples, which was not discussed in \cite{2016Flow}. 
In practical workload, we discover that none of PRPD (\resp Flow-Join), PnR and the elegant parallel grace hash join~\cite{DBLP:conf/icde/KitsuregawaTN92} (referred to as GraHJ) fits to all workloads. The observation motivates us to develop a self-adaptive solution, in which a cost-based decider is designed to choose the optimal algorithm depending on the given workloads.

\begin{table}[t]
  \caption{Notations.}
  \label{tab:tb1}
  \footnotesize
  \begin{tabular}{cp{5cm}l}
    \toprule
    Symbol & Description\\
    \midrule
     $R$, $S$ & probe table and build table \\
     $R^i$, $S^i$ & a share of $R, S$ at node $i$ \\ 
     $|R|$, $|S|$ & amount of the tuples in $R, S$\\
     $N$ & number of computation nodes enrolled in \textsf{Dist-HJ}\\
     $p$ & skew threshold\\
     $\rho_a^p(R)$, $\rho_b^p(S)$ & skewed value sets of $R, S$\\
     $\rho_{(C)}(R,S)$ & set of the tuples in $R, S$ who is complete skewed\\
     $\rho_{(C+)}(R,S)$ & set of the tuples in $R, S$ who is left-dominated complete skew\\
     $\rho_{(C-)}(R,S)$ & set of the tuples in $R, S$ who is right-dominated complete skew\\
     ${R}_1$ & set of all the non-skewed tuples in $R$ \\
     ${R}_2$ & set of the \textit{partial skew} tuples in $R$ \\
     ${R}_3$ & set of the tuples who belongs to $\rho_{(C+)}(R,S)$ \\
     ${R}_4$ & set of the tuples who belongs to $\rho_{(C-)}(R,S)$ \\
     ${S}_2$ & set of tuples $t\in S$ who are \textit{partial skewed} in $R$ \\
     ${S}_3$ & set of the tuples $t\in S$ and $t.a\in\rho_{(C+)}(R,S)$ \\
     ${S}_4$ & set of the tuples $t\in S$ and $t.a\in\rho_{(C-)}(R,S)$ \\
     ${S}_1$ & set of the tuples $t\in S \setminus {S}_2 \setminus {S}_3 - {S}_4$ \\
     $\mathcal{R}_{hash}$ & tuples are \textit{hash-distributed} in $R$ \\
     $\mathcal{R}_{loc}$ & tuples are \textit{kept local} in $R$ \\
     $\mathcal{R}_{rand}$ & tuples are \textit{randomly distributed} in $R$ \\
     $\mathcal{R}_{repl}$ & tuples are \textit{replicated and sent
to all nodes} in $R$ \\
     $\mathcal{S}_{hash}$ & tuples are \textit{hash-distributed} in $S$ \\
     $\mathcal{S}_{rand}$ & tuples are \textit{randomly distributed} in $S$ \\
     $\mathcal{S}_{repl}$ & tuples are \textit{replicated and sent
to all nodes} in $S$ \\
     $sel(R,x)$, $sel(S,x)$ &  the selectivity of $x$ in $R$ or $S$ \\
     $M_h(x)$ & the cost of value $x$ if hash-redistributed  \\
     $M_h(R,x)$, $M_h(S,x)$ & the cost of value $x$ in $R$ or $S$ if hash-redistributed  \\
     $M_l(R,x)$, $M_l(S,x)$ & the cost of value $x$ in $R$ or $S$ if kept local \\
     $M_r(R,x)$, $M_r(S,x)$ & the cost of value $x$ in $R$ or $S$ if randomly distributed \\
     $Q(R,x)$, $Q(S,x)$ &  the number of tuples with value $x$ in $R.a$ or $S.b$ \\
    \bottomrule
  \end{tabular}
\end{table}

\section{Partition and Replication}\label{sec3}
In this section we propose a novel \textsf{Dist-HJ} solution towards skewed workload, namely Partition and Redistribution (PnR). In PnR, we focus on the standard scenario of \textsf{Hash Join}, where $R$ (probe table) considerably outnumbers $S$ (build table). For ease of discussion, we present a group of definitions first (key notations are summarized in~\autoref{tab:tb1}).
\begin{definition}[Skew]\label{def:skew}
Given a table $R$ and a skew threshold $p$, a value $x$ is \textit{Skewed} in $R.a$ ($a$ is a column in $R$) if $|\sigma_{R.a=x}(R)|\ge p|R|$. For ease of discussion, let $\rho_a^p(R)=\{x||\sigma_{R.a=x}(R)|\ge p|R|\}$ (denoted as $\rho(R)$ for brevity) refer to the set of these $x$.
\end{definition}
\begin{definition}[Complete Skew]\label{def:cskew}
Given that we are performing an equi-join over $R$ and another table $S$, \ie $R\underset{R.a=S.b}{\bowtie }S$, if $x$ is skewed in both $R.a$ and $S.b$, \ie $x\in\rho(R)\cap\rho(S)$, it is referred to as \textbf{Complete Skewed}, denoted as $\rho_{(C)}(R,S)$. 
Specifically, given $x\in \rho_{(C)}(R,S)$, if $|\sigma_{R.a=x}(R)|>|\sigma_{S.b=x}(S)|$, we refer to it as \textbf{Left}(\resp \textbf{Right})\textbf{-dominated Complete Skew}, denoted as $\rho_{(C+)}(R,S)$ (\resp $\rho_{(C-)}(R,S)$).
\end{definition}
\begin{definition}[Partial Skew]\label{def:pskew}
Given that we are performing $R\underset{R.a=S.b}{\bowtie }S$, if $x$ is skewed in $R.a$ but not in $S.b$, \ie $x\in\rho(R)\setminus\rho(S)$, it is referred to as \textbf{Partial Skewed} in $R$, denoted as $\rho_{(P)}(R,S)$.  
\end{definition}
\vspace{-1ex}\subsection{Limitations of PRPD}
According to the assumption of PRPD, when $R^i$ and $S^i$ are balanced over all computation nodes of the cluster, the number of skewed value on each node are also assumed to be balanced accordingly. However, when we introduce PRPD to KaiwuDB and replace the default GraHJ solution, the following problems are observed:
\begin{enumerate}
\item Suppose that we are preforming $R\underset{R.a=S.b}{\bowtie }S$, when a value $x$ of the join columns is \textit{partial skew} in $S.b$, \ie $x\in\rho(S)\setminus\rho(R)$, the tuples of $x$ in $R$ (\ie $\{t|t\in R, t.a=x\}$) will be broadcast and those in $S$ (\ie $\{t|t\in S, t.b=x\}$) will be kept local, according to PRPD. As $|R|\gg|S|$, it is highly possible that $|\sigma_{R.a=x}(R)| \gg |\sigma_{S.b=x}(S)|$, in which case broadcasting the tuples \wrt $x$ in $R$ will inevitably introduce too much computational burden.  
\item In PRPD, computation workload of each node is unbalanced because of the significant difference between the volume of $R^i$ and $S^i$. In KaiwuDB, each $R^i$ or $S^i$ reads data through a \textsf{Tablereader} operator, the number of records read by which are not always identical, further enlarging the imbalance among $R^i$ (\resp $S^i$).
\item In PRPD-u, the random distribution of all skewed values introduces unnecessary data transmission task. At the same time, the corresponding values on the other table also need to be broadcast, which inevitably results in significant fluctuation in the aspect of transmission burden.
\end{enumerate}
To address the above issues, we present a new solution of \textsf{Dist-HJ}, namely PnR. In PnR, we focus on the skewed value of the probe table, as it is the key to the above problem (1). We partition the skewed value (\ie $\rho(R)$) in the big table (\eg $R$) into two parts: 1) for those \textit{complete skew} values, \ie $\rho(R)\cap \rho(S)$, we divide them equally, and the corresponding records of the small table, \ie $\sigma_{S.b\in\rho(R)\cap \rho(S)}(S)$, is broadcast; 2) for the \textit{partial skew} values in $R.a$, \ie $\rho(R)\setminus \rho(S)$, we follow the PRPD strategy and keep them locally at node $i$. We advocate and will show later that this will address the above problems (2) and (3). 

\vspace{-1ex}\subsection{Description of PnR}\label{section:PnR}
\begin{figure}
  \centering
  \includegraphics[width=\linewidth]{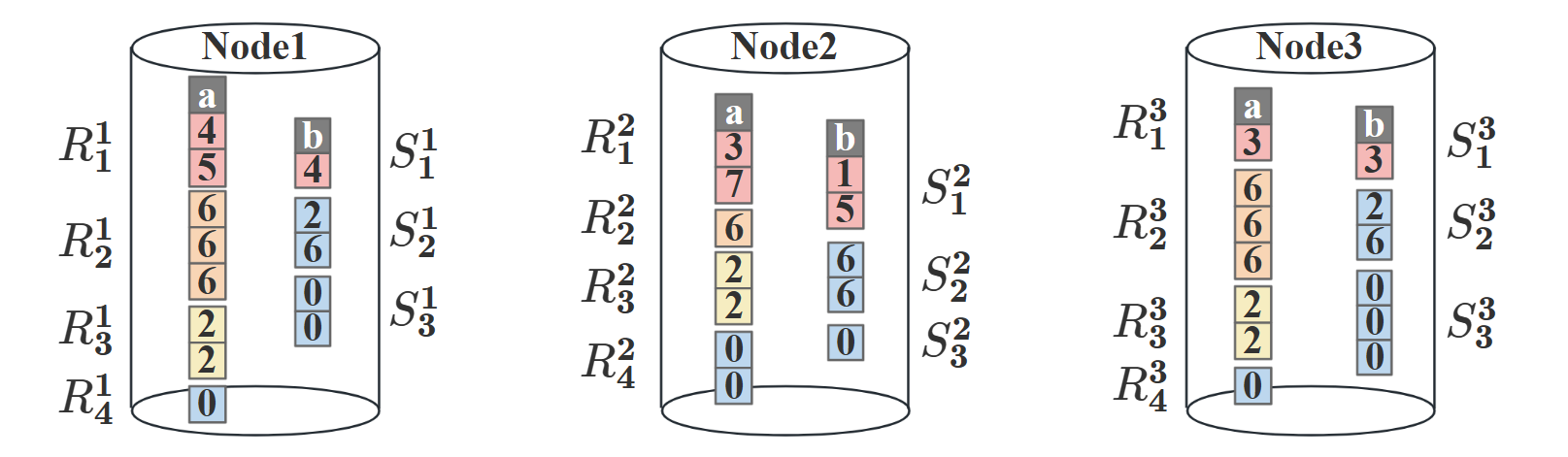}
  \vspace{-3ex}\caption{Partitions of $R$ and $S$ on a 3-node cluster, where "6" is a \textit{complete skew} value, "2" and "0" are the \textit{partial skew} values in $R$ and $S$, respectively. The hash function used in the figure is $h(x)=(x+1)\mbox{ mod 3}$. }
  \label{fig:Redistribution}
\end{figure}
\begin{figure}
  \centering
  \includegraphics[width=\linewidth]{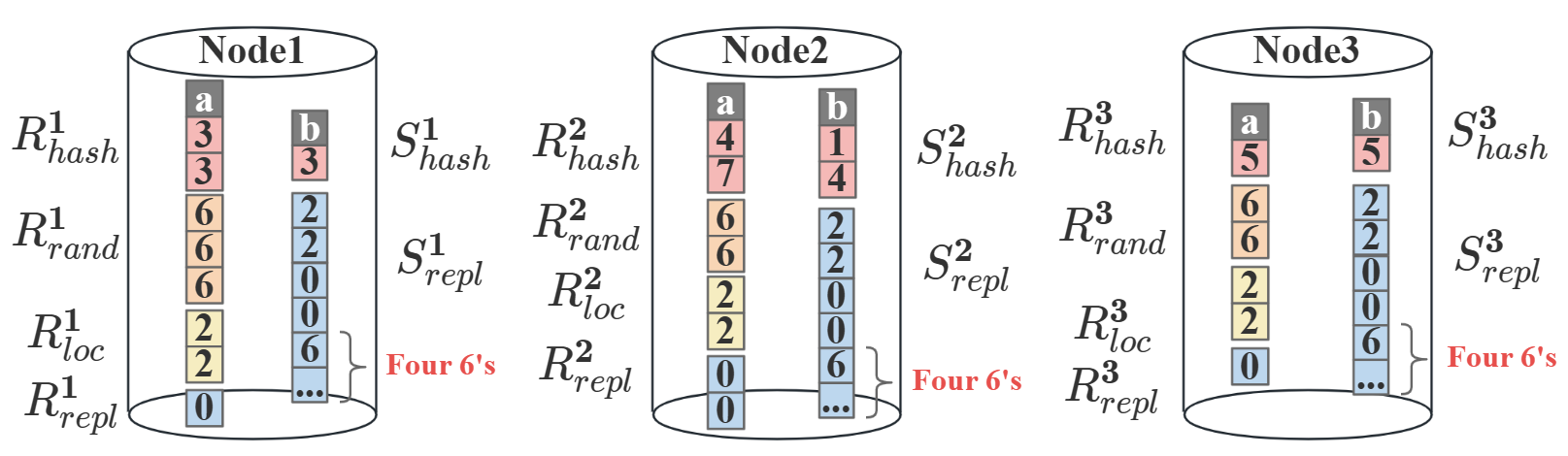}
  \vspace{-2ex}\caption{Data placement after redistribution strategy of PnR, where the hash function is $h(x)=(x+1)\mbox{ mod 3}$.}
  \label{fig:After_redistribution}
\end{figure}
Suppose that the optimizer chooses \textsf{Hash Join} to preform $R\underset{R.a=S.b}{\bowtie }S$ ($|R|>|S|$) in the distributed \textsc{rdbms}. As frequency is nowadays collected in many \textsc{rdbms}s~\cite{2020CockroachDB,2020Ti,lyu2021greenplum,DBLP:conf/icde/CaoLHLZHSZWWLCF22}, the statistics on skewed values of both tables are known \textit{apriori}.
In PnR, the \textit{complete skew} values, \ie $\rho(R)\cap \rho(S)$, are further divided into left-dominated and right-dominated ones according to Definition~\autoref{def:cskew}, namely $\rho_{(C+)}(R,S)$ and $\rho_{(C-)}(R,S)$.  

Given the above, we shall describe the key steps in PnR as follows.

\btitle{Step 1, Redistribution.} 
Similar to PRPD and GraHJ, before the \textsf{Dist-HJ} is executed, each node $i$ in the cluster obtains a share of $R$, say $R^i$, from the output of the precedent operator. Different with PRPD, which partitions $R^i$ into three disjoint sets, PnR selects to partition it into four parts as follows. 
\begin{description}
\item [${R}^{i}_1$:] This part contains all the non-skewed tuples $t$, \ie $t.a\in{R}^ {i}.a\setminus\rho({R}^{i})$. They are \textit{\textbf{hash-redistributed}} (identical to GraHJ).
\item [${R}^{i}_2$:] This part are the tuples $t$ whose entries in $a$ are \textit{partial skew} in $R$. They are \textit{\textbf{kept local}} at node $i$.
\item [${R}^{i}_3$:] The third set contains the tuples $t\in R^i$, where $t.a\in\rho_{(C+)}(R,S)$. They are \textit{\textbf{randomly distributed}}.
\item [${R}^{i}_4$:] It contains the tuples $t\in R^i$, where $t.a\in\rho_{(C-)}(R,S)$. They are \textit{\textbf{broadcast}} to all nodes.
\end{description}  
Given the above partitions of $R^i$ and their corresponding redistributing strategies, each node $i$ will receive four parts accordingly, referred to as $\mathcal{R}^{i}_{hash}$, $\mathcal{R}^i_{loc}$, $\mathcal{R}^{i}_{rand}$ and $\mathcal{R}^{i}_{repl}$, respectively. For instance, $\mathcal{R}^{i}_{hash}$ contains the tuples $t$ that $hash(t.a)\ mod\ N = i$ in ${R}^{i}_{1}$ for all $i=1,\ldots,N$ ; $\mathcal{R}^i_{loc}$ are identical to ${R}^{i}_{2}$, as it is kept local at $i$; $\mathcal{R}^{i}_{rand}$ contains a random share of the tuples $t$ uniformly sampled from $\cup_{i=1}^N{R}^{i}_3$; $\mathcal{R}^{i}_{repl}$ is identical to $\cup_{i=1}^N{R}^{i}_4$, as any tuple of it is broadcast to all nodes.

Aside from the above strategies with respect to $R^i$, we need to divide $S^i$ into disjoint sets accordingly, resulting in $S^i_1, S^i_2, S^i_3, S^i_4$. Differently, the redistribution strategy of them should depend on how their counterpart in $R$, \ie those tuples in $R$ satisfying the join condition, are distributed. When the tuples $\{t|t.a=x, t\in R\}$ are distributed to multiple nodes, the corresponding value on $S$, \ie $\{s|s.b=x, s\in S\}$, must be broadcast to all nodes, so as to ensure the correctness of the join results. Driven by that, we shall show in detail the redistribution strategy for $S^i_1, S^i_2, S^i_3, S^i_4$ as follows (a running example can be found in Figure~\ref{fig:Redistribution}):

\begin{description}
\item [${S}^{i}_1$:] This part consists of tuples $t\in S^i\setminus {S}^{i}_2\setminus {S}^{i}_3\setminus {S}^{i}_4$. They are \textit{\textbf{hash-redistribute}d}.
\item [${S}^{i}_2$:] This part contains tuples $t\in S^i$ whose entries in $a$ are \textit{partial skew} in $R$. They are \textit{\textbf{broadcast}} to all nodes.
\item [${S}^{i}_3$:] The third set contains the tuples $t\in S^i$, where $t.a\in\rho_{(C+)}(R,S)$. They are \textit{\textbf{broadcast}} to all nodes.
\item [${S}^{i}_4$:] It contains the tuples $t\in S^i$, where $t.a\in\rho_{(C-)}(R,S)$. They are \textit{\textbf{randomly distributed}}.
\end{description} 

Finally, each node $i$ will get three shares of $S$, namely $\mathcal{S}^{i}_{rand}$, $\mathcal{S}^{i}_{repl}$ and $\mathcal{S}^{i}_{hash}$. In particular, $\mathcal{S}^{i}_{rand}$ are those obtained by random distributing $S^i_4$ obtained from all nodes, that is, it contains tuples $t$ such that $t.a\in\rho_{(C-)}(R,S)$; $\mathcal{S}^{i}_{repl}$ are tuples obtained after \textit{broadcasting} $S^i_2\cup S^i_3$; $\mathcal{S}^{i}_{hash}$ are those tuples \textit{hash-redistributed} from all ${S^i_1}$ ($i=1,\ldots,N$). Figure~\ref{fig:After_redistribution} shows an example after \textbf{Step 1}.

\btitle{Step 2, Join.} 
After redistribution, each node receives four sets of tuples from $R$ and three from $S$. Afterwards, each node has to perform the following joins locally to get the correct results:
$$\mathcal{R}^{i}_{hash}\underset{a=b}\bowtie{\mathcal{S}^{i}_{hash}}, \mathcal{R}^{i}_{loc}\underset{a=b}\bowtie{\mathcal{S}^{i}_{repl}}$$
$$\mathcal{R}^{i}_{rand}\underset{a=b}\bowtie{\mathcal{S}^{i}_{repl}}, \mathcal{R}^{i}_{repl}\underset{a=b}\bowtie{\mathcal{S}^{i}_{rand}}$$
The first join implements the \textsf{Hash Join} over the non skewed tuples in $R$ and $S$. In the second and third join, although the tuples in $\mathcal{R}^{i}_{loc}$ and $\mathcal{R}^{i}_{rand}$ are distributed on $i$-th node, $S$ broadcasts the join-able tuples so that each node receives the same $\mathcal{S}^{i}_{repl}$. Therefore, valid results can never escape from the output of the second and third joins. Similar to the above, there is no missing result in the fourth join. Hence, the above four joins contain all valid results of $R\underset{R.a=S.b}{\bowtie }S$.

\btitle{Step 3, Union.} The final result is obtained by the union of the results from four joins performed all node. 

In all, the pseudocode of PnR is outlined in Algorithm~\ref{alg:PnR}. Lines 4 to 11 describes the Redistribution phase. Line 4 to Line 5 initializes $R^i, S^i$, respectively. Line 7 divides $R^i, S^i$ into multiple partitions accordingly. Lines 8 to 11 implements different redistribution policies for different partitions. Lines 13 to 21 refers to the Join phase. Line 23 integrates results from all nodes. 

\begin{algorithm}
\renewcommand{\algorithmicrequire}{\textbf{Input:}} 
\renewcommand{\algorithmicensure}{\textbf{Output:}}
\caption{PnR Algorithm}\label{alg:PnR}
\begin{algorithmic}[1]
\Require $R$; $S$; The skewed value sets $\rho(R),\rho(S)$; 
\Ensure The result set of PnR $result$


\State $result \gets []$
\For{node $i$ in $n$ nodes}
    \State // Step 1, Redistribution.
    \State $R^i \gets R$ on the $i$-th node
    \State $S^i \gets S$ on the $i$-th node
    \State // \Call{Partition}{$R^i, S^i$} refers to Step 1 of Section \ref{section:PnR}
    \State $R^i_1,R^i_2,R^i_3,R^i_4,S^i_1,S^i_2,S^i_3,S^i_4 \gets$ \Call{Partition}{$R^i, S^i$} 
    \State hash-redistribute $R^i_1,S^i_1$
    \State kept local $R^i_2$
    \State randomly distribute $R^i_3,S^i_4$
    \State broadcast $R^i_4,S^i_2,S^i_3$
    
    \State // Step 2, Join.
    \State $\mathcal{R}^{i}_{hash} \gets \{t | hash(t.a) \bmod N = i, t \in \cup^{N}_{i=1} R^i_1 \}$
    \State $\mathcal{R}^i_{loc} \gets R^i_2$ 
    \State $\mathcal{R}^{i}_{rand} \gets$ tuples $t$ uniformly sampled from $\cup_{i=1}^N{R}^{i}_3$
    \State $\mathcal{R}^{i}_{repl} \gets \cup_{i=1}^N{R}^{i}_4$
    \State $\mathcal{S}^{i}_{rand} \gets$ tuples $t$ uniformly sampled from $\cup_{i=1}^N{S}^{i}_4$
    \State $\mathcal{S}^{i}_{repl} \gets \cup_{i=1}^N (S^i_2\cup S^i_3)$ 
    \State $\mathcal{S}^{i}_{hash} \gets \{t | hash(t.a) \bmod N = i, t \in \cup^{N}_{i=1} S^i_1 \}$
    \State $result^i = (\mathcal{R}^{i}_{hash}\underset{a=b}\bowtie{\mathcal{S}^{i}_{hash}}) \cup (\mathcal{R}^{i}_{loc}\underset{a=b}\bowtie{\mathcal{S}^{i}_{repl}})$
    \State $result^i = result^i \cup (\mathcal{R}^{i}_{rand}\underset{a=b}\bowtie{\mathcal{S}^{i}_{repl}}) \cup (\mathcal{R}^{i}_{repl}\underset{a=b}\bowtie{\mathcal{S}^{i}_{rand}})$
    
    \State // Step 3, Union.
    \State $result = result \cup result^i$
\EndFor
\State \textbf{return} $result$
\end{algorithmic}
\end{algorithm}

\vspace{-2ex}\subsection{Limitations}\label{sec33}
In the PnR algorithm, it is assumed that the volume of data in the probe table significantly exceeds that in the build table, so we only deal with the skewed values in the probe table and ignore those in the build one. Although this assumption is popular and consistent with most practical scenarios, there exist some exceptions as follows:
\begin{enumerate}
\item $|R|\approx|S|$, $\rho(R)=\emptyset$. In this case, $\rho(S)$ is not processed and PnR degenerates into GraHJ. On the contrary, PRPD can perform well because $\rho(S)$ is kept local on each node.
\item $|R|\approx|S|$, $\rho(R)\cap\rho(S)=\emptyset$. Similarly, PnR still does not deal with $\rho(S)$. In some cases, this will lead to extremly uneven computation at each node after redistribution.
\end{enumerate}
In general, PnR does not care about partial skewed tuples of $S$, which may lead to the unbalanced computation burden of each node after redistribution. As the volume of $S$ and $R$ become close to each other, the imbalance will be more significant.

\section{A Self-Adaptive Solution in K\lowercase{aiwu}DB}\label{sec:4}
Given the three representative solutions towards \textsf{Dist-HJ}, namely PRPD, PnR and GraHJ, the above theoretical study implies that there is not a single algorithm that wins in all the scenarios. Generally, PRPD is optimal when the join table are evenly partitioned and the skewed data are evenly distributed. PnR takes into account both the network traffic and the skewness of the join results, and performs well in most of the skewed workloads, except for the scenarios discussed in Section~\ref{sec33}. GraHJ is optimal when a large number of skewed tuples are assigned to the gateway node (where the query is initially received) and all results need to be gathered accordingly. Given the above findings, in order to better address the skew issue of \textsf{Dist-HJ} within KaiwuDB, we are motivated to develop a mechanism that adaptively finds the most optimal solution at runtime given a particular workloads. In this part, we shall first give some concrete examples for the phenomenon, and present to model different solutions, \ie PRPD, PnR and GraHJ, as 
sub-operators attached to \textsf{Hash Join}. In addition, we propose a cost-based sub-operator dispatcher that is native embedded into the cost model and optimizer of KaiwuDB.

\begin{figure*}[t]
  \centering
  \includegraphics[width=\linewidth]{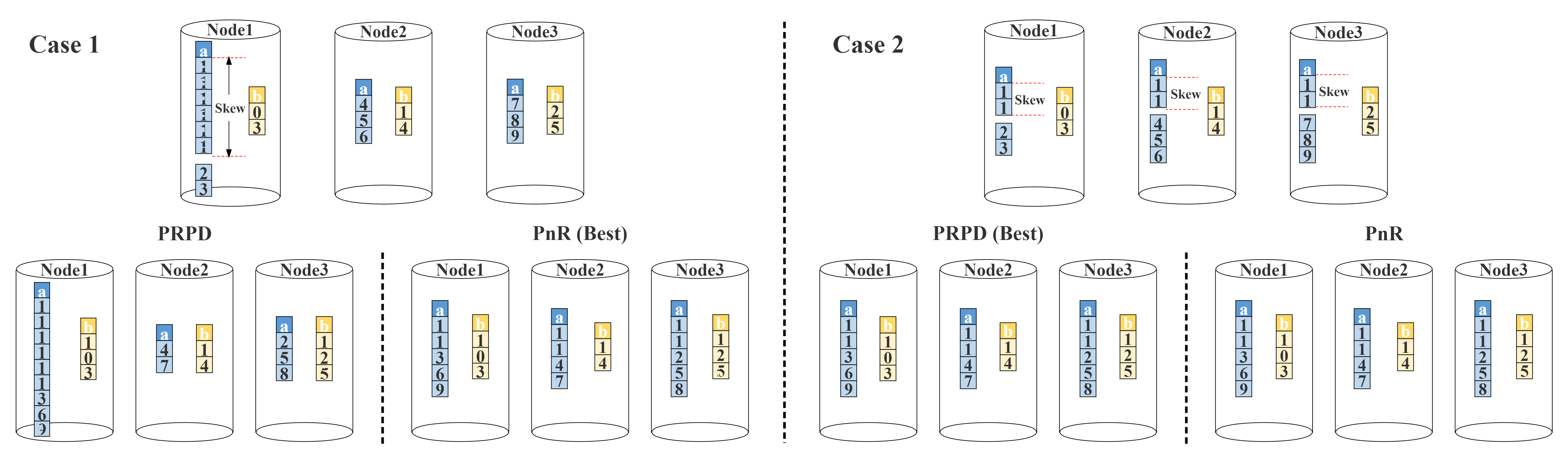}
\vspace{-3ex}\caption{A running example showing the difference of PRPD, PnR under different skew scenarios}
  \label{fig:change}
\end{figure*} 
\vspace{-2ex}\subsection{One Size Cannot Fit All}
To illustrate the problem that no single solution wins all scenarios, we shall investigate three representative cases. In each case, we shall study the time complexity (in term of the number of tuples to be processed) of different solutions and show also our corresponding empirical results afterwards. For simplicity, suppose $R$ contains a single skewed value $x$, while $sel(R,x)$ and $sel(S,x)$ represent the selectivity of $x$ in $R$ and $S$, respectively.

\btitle{Case 1. $sel(S,x)> 0$  and $sel(R,x)= 0$.} According to the followings, PRPD wins in this case.
    \begin{description}
        \item[GraHJ.] The number of tuples to be processed by the slowest node can be estimated as $|S|sel(S,x) + \frac{|R|}{N}$.
        \item[PnR.] It degenerates to GraHJ, and its complexity is identical to that of GraHJ.
        \item[PRPD.] The complexity of the slowest node can be estimated as $\max{\{|S^i|\}}sel(S,x) + \frac{|R|}{N}$. Obviously, $\max{\{|S^i|\}}sel(S,x)<|S|sel(S,x)$.
    \end{description}
\btitle{Case 2. $sel(S,x)> 0$  and $sel(R,x)> 0$.} According to the followings, PnR wins in this case.
    \begin{description}
        \item[GraHJ.] The time complexity of the slowest node can be estimated as $|S|sel(S,x) + |R|sel(R,x) + |S|sel(S,x)\cdot|R|sel(R,x)$, where $|S|sel(S,x)\cdot|R|sel(R,x)$ is the materialization time.
        \item[PnR.] The complexity for the slowest node is $\max{\{|S^i|\}}sel(S,x) + \frac{|R|sel(R,x)}{N} + \max{\{|S^i|\}}sel(S,x) \cdot \frac{|R|sel(R,x)}{N}$.
        \item[PRPD.] The complexity can be estimated as $\max{\{|S^i|\}}sel(S,x) + |R^i|sel(R,x) + |{S^i}|sel(S,x) \cdot |R^i|sel(R,x)|$.
    \end{description}
\btitle{Other cases.} In both the above cases, we assume that the skewed tuples stay at the gateway   node. Otherwise, a large number of results are generated in the gateway node. In that case, GraHJ algorithm will greatly reduce the network overhead and wins.

\autoref{fig:change} shows the redistribution scheme of both PnR and PRPD under different data distribution scenarios. For simplicity, other columns of table $R$ and $S$ are not shown but the join ones, \ie $R.a$ and $S.b$. In both scenarios, there is a partial skew value, \ie $1$. In Case 1, the skew value 1 are all located on \textit{Node1}. Following the strategy of PRPD, these values have to be kept on \textit{Node1}. In comparison, according to PnR, the entries of the partial skew value, \ie $1$, have to be evenly redistributed over the cluster, in which case the workload are more balanced than PRPD. In Case 2, the partial skew value, \ie $1$, is already evenly distributed over the three nodes before the join happens. In both PRPD and PnR, these corresponding entries will be kept local at the original node. Thus, the workload is balanced in both PRPD and PnR, except for the difference that the skew value 1 will not introduce extra network overhead in PRPD.

\begin{figure}[t]
  \centering
  \includegraphics[width=.75\linewidth]{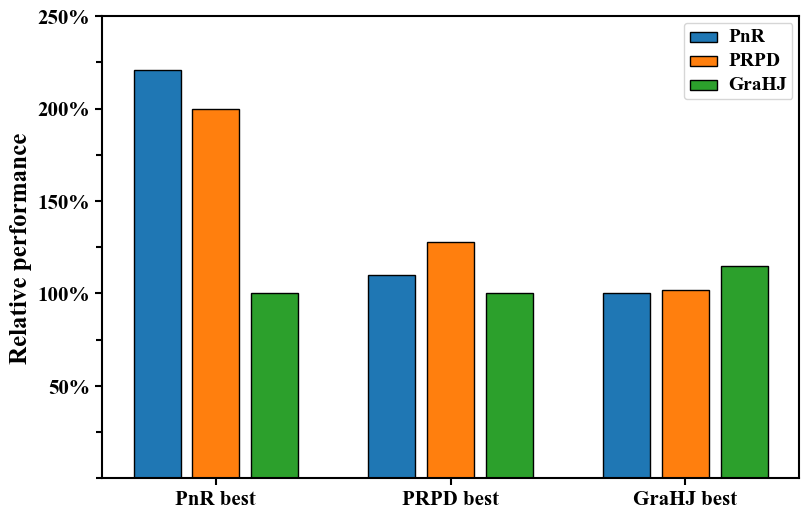}
  \vspace{-3ex}\caption{Performance comparison of three algorithms in different scenarios}
  \label{fig:comparison}\vspace{-4ex}
\end{figure}

Besides the above theoretical analysis and running example, we additionally conduct exhaustive experiments accordingly. We implemented PRPD, PnR and GraHJ in KaiwuDB, and reproduced the above three scenarios, using a 3-nodes cluster platform with Intel Xeon gold 6240, 64GB memory on each single node. The experimental results are shown in \autoref{fig:comparison}. It shows that in the first case, PnR is $10\%$ and $121\%$ faster than PRPD and GraHJ. In the second case, PRPD is about $16\%$ and $28\%$ faster than PnR and GraHJ, respectively. In the last case discussed above, GraHJ is $15\%$ and $13\%$ faster than PnR and PRPD.

\vspace{-2ex}\subsection{Self-Adaptive Strategy}\label{ssec42}
Since any single algorithm of GraHJ, PRPD and PnR fails to fit all workloads, hereby we propose a cost-based dispatcher to automatically choose the best solution depending on the characteristic of the target workload. As known to all, Cost-Based Optimizer (CBO) is adopted in the major \textsc{rdbms}s to find the query execution plan (\textsc{qep}) given a \textsc{sql}. Table statistics are used to estimate the cost of each candidate plan, and CBO chooses the plan with the lowest cost. For instance, in KaiwuDB, \textsf{Dist-HJ} maybe adopted by a \textsc{qep} as other alternatives of join, \ie \textsf{Nested Loop, Sort Merge}, may lead to larger cost than GraHJ. Taking into account our insights in the last subsection, considering that GraHJ is one of a series of alternative implementations \wrt \textsf{Dist-HJ}, the estimate of the cost of \textsf{Dist-HJ} operator can be no long solely based that of GraHJ, but on the minimum among the three alternatives, \ie PRPD, PnR, GraHJ. Driven by that, we propose to treat the different implementations of \textsf{Dist-HJ} as sub-operators, each of which will contribute to the cost estimate of \textsf{Dist-HJ} independently. Obviously, the sub-operator that exhibits the lowest cost will be adopted by the \textsc{qep}. In order to estimate the cost for each sub-operator, we need the following information:
\begin{itemize}
\item The distribution of \textsf{Tablereader}s for both $R$ and $S$, which implies the nodes participating the join and how much data are accessed at each individual node, \ie $R^i, S^i$.  
\item The frequency statistics of the join columns, which implies the skew values $x$.
\end{itemize}
We generalize the execution of \textsf{Dist-HJ} into three phases, namely redistribution, join and merge. Each phase corresponds to a cost, and our cost-based dispatcher compares the three costs to decide which specific algorithm to choose. Next, we shall illustrate each of these phases in detail:

\noindent\btitle{Redistribution.} In \textsf{Hash Join}, the smaller table $S$ are used to build the hashtable, and the bigger table $R$ is probed accordingly. As long as all of the tuples of $S$ have been redistributed, can the probing phase start. The probing phase ends when the last batch of $R$ finishes probing. Redistribution affects the start and end time of the join phase, and accounts for the largest share within the execution time of \textsf{Dist-HJ}. Different redistribution plans result in different amount of data to be transmitted over network. Notably, among all the choices (\ie PRPD, PnR, GraHJ), non-skewed tuples are redistributed in the same way; the number of skewed tuples interchanged over network differs much. Therefore, we use the volume of skewed tuples interchanged over network to measure the cost for the redistribution phase for each sub-operator \wrt \textsf{Dist-HJ}\footnote{As the time cost for this data exchange varies largely \wrt network devices, we select to compare only the volume of data, which direct reflects the time cost once the network device is fixed.}. \eat{Note that \textsf{Tablereader}s are the executors of the redistribution, the costs are calculated per \textsf{Tablereader}, and are added up to be the cost the redistribution of \textsf{Dist-HJ}. }

\noindent\btitle{Join.} 
Each node performs hash join individually once the tuples are redistributed. We divide the cost of this phase into three parts, building, probing and materialization. The smaller table is scanned to build the hashtable, and the bigger table is probed to find the matched results.  After the two phases, results produced by the join are materialized for the execution of the subsequent processors. The cost of materialization depends on the volume of the join results. The cost of the three parts are summed up then. Note that hash joins are performed on all $N$ nodes in parallel, we use the maximum cost among all nodes as the cost for this phase. 

\noindent\btitle{Merge.} 
Merge refers to the process when local join results at each node are further transmitted across the cluster, which varies with respect to the eventual distribution of the join results and the subsequent execution plan. The volume of join results on nodes differs from each other due to the precedent redistribution and different join selectivities at each nodes. No matter which algorithm the optimizer chooses, the total number of the join results is fixed. Depending on the subsequent operator along the \textsc{qep}, we classify it into two types. In the first type, the join results are gathered as the final result by one gateway node. In order to achieve that, the local join results at each node are materialized and sent to the gateway node. Note that the results on gateway node do not need to be transmitted over the network. In the second type, the join results are kept local and processed as intermediate products of the subsequent operator. For example, considering the query,
\begin{lstlisting}[language=SQL,keywordstyle=\color{blue},xleftmargin=.025\textwidth,xrightmargin=.05\textwidth
]
SELECT COUNT(*) FROM R,S WHERE R.a=S.b;  
\end{lstlisting}    
the join results are aggregated and do not need to 
be entirely sent to the gateway node. Note that the subsequent operators are not parts of \textsf{Dist-HJ}, and the costs of which should not be taken into account for the Merge step. The costs are calculated per node and summed up then. In particular, for the first type of subsequent operator, the cost of Merge depends on the total number of join results on the non-gateway nodes; for the second type of subsequent operator, the cost of Merge is $0$, as the join results are kept locally. 
Therefore, the cost of Merge is query-dependent. 

Indeed, the three sub-operators perform the same for non-skewed tuples, hence we only need to study the difference in the cost produced by those skewed values.
 
\subsection{Cost model of the sub-operators}
According to the above study, the cost of the Join and Merge steps depend on the redistribution plan. Therefore, before diving into the cost model for the sub-operators, we study the characteristics of each redistribution strategy. 
  
Generally, among all the sub-operators, tuples can only be redistributed in four ways, 
\begin{itemize}
    \item [i)] hash redistributed;
    \item [ii)] kept local;
    \item [iii)] randomly redistributed;
    \item [iv)] broadcast.
\end{itemize}
Once the redistribution strategy for tuples with entry $x$ on the join column in one table is fixed, the strategy for the correlated tuples (with entry $x$ on the join column) in the other table is decided, according to our study in~\autoref{sec3}. 

Given a value $x$, using the first approach, when tuples $t$ with value $x$ in one table are hash redistributed, the correlated tuples in the other table are also hash redistributed. For instance, suppose $t\in R$ and $t.a=x$, once $t$ is hash redistributed to some node $i$, those tuples $t^\prime\in S$ and $t^\prime.b=x$ have to be redistributed to the same node $i$, in order to guarantee the correctness in the results.

Using the second approach, tuples with value $x$ are kept local in one table, and correspondingly tuples in the other table are broadcast. For instance, suppose $t\in R$ and $t.a=x$, once $t$ is kept local, those tuples $t^\prime\in S$ and $t^\prime.b=x$ can be broadcast.

Under the third approach, tuples with value $x$ are randomly redistributed to all nodes, and correspondingly tuples in the other table are broadcast. For instance, suppose $t\in R$ and $t.a=x$, once $t$ is randomly redistributed, those tuples $t^\prime\in S$ and $t^\prime.b=x$ should be broadcast.

For ease of discussion, we denote the redistribution operation i), ii) and iii) over tuples (in a table $R$) whose entries on the join column is $x$ as $M_h(R,x)$, $M_l(R,x)$ and $M_r(R,x)$, respectively. For instance, $M_l(R,x)$ denotes the method using which tuples $t\in\{t|t\in R,t.a=x\}$ is kept local, according to ii), the corresponding tuples in $S$ are broadcast. Notably, given a join task over $R$ and $S$, tuples may follow one of the above redistribution strategy, and an overall redistribution plan for the join task can be described as a set of redistribution methods, each of which applies for a subset of the tuples. For example, redistribution plan of GraHJ can be described as $\forall x\in R.a, M_h(R,x)$ and $\forall x\in S.b, M_h(S,x)$. 

For convenience, we use $Q(R,x)$ and $Q(S,x)$ to denote the number of tuples with value $x$ in $R.a$ and $S.b$, respectively. $Q(R^i,x)$ and $Q(S^i,x)$ respectively refers to the number of tuples of value $x$ in $R$ and $S$ on the $i$-th node. We shall study the cost of each sub-operator in terms of three \textsf{Dist-HJ} phases, \ie Redistribution, Join and Merge, respectively.

\btitle{Redistribution.}
As discussed in the above, there are three types of redistribution strategies, \ie $M_h, M_l, M_r$. 

For $M_h$, assuming that tuples $t$ with $t.a = x$ are assigned to the $i$-th node, \ie $hash(x)\bmod{N}=i$. Tuples $t$ with $t.a = x$ from all \textsf{Tablereader}s are hash redistributed except those on the $i$-th node. The volume for the redistributed tuples with value $x$ over the network is denoted as $|{M_h}(x)|=Q(R,x)+Q(S,x)-\frac{Q(R,x)+Q(S,x)}{N}$, which directly reflects the cost of $M_h$. 

For $M_l$, when tuples with value $x$ from $R^i$ are kept local, the tuples with value $x$ from $S^i$ is broadcast, in this case tuples with a volume $Q^i(S,x)(N-1)$ are sent over the network. The total number of the tuples with value $x$ sent over the network is $|M_{l}(R,x)|=Q(S,x)(N-1)$, and $|M_{l}(S, x)|=Q(R,x)(N-1)$ when tuples with value $x$ from $S^i$ are kept local. 

For $M_r$, assuming that the tuples with value $x$ from $R^i$ are randomly assigned to all nodes with uniform probability of $\frac{1}{N}$. Consequently, the $i$-th node will send $Q^i(R,x)\frac{(N-1)}{N}$ tuples over network, and the tuples with value $x$ from $S^i$ is broadcast, whose volume can be modeled as $Q^i(S,x)(N-1)$. In total, under $M_r$, the volume of tuples sent over the network is $|M_{r}(R, x)|=Q(R,x)\frac{(N-1)}{N}+Q(S,x)(N-1)$, and $|M_{r}(S, x)|=Q(S,x)\frac{(N-1)}{N}+Q(R,x)(N-1)$ when tuples with value $x$ from $S^i$ is randomly redistributed. 

Note that the costs of each redistribution strategy for a single joined value, \eg $x$, are modeled in pairs, containing the number of redistributed tuples of both tables. These strategies are adopted in hybrid manner for each sub-operator, \ie GraHJ, PRPD and PnR, the details of which are listed in the follows. 
\begin{itemize}
    \item [A)] GraHJ.
    In GraHJ, skewed tuples are hash redistributed. The redistribution cost of GraHJ can be estimated as 
\begin{equation}
\begin{split}
    Re_{GraHJ}=\sum_{x\in{\rho_a^p(R)\bigcup{\rho_b^p(S)}}}{|M_{h}(x)|}
\end{split}
\end{equation}

    \item [B)] PRPD. Here, the skewed tuples are always kept local, so the redistribution cost is
\begin{equation}
\begin{split}
    Re_{PRPD}=\sum_{x\in{\rho_a^p(R)}}{|M_{l}(R,x)|} + \sum_{x\in{\rho_b^p(S)}}{|M_{l}(S,x)|}
\end{split}
\end{equation}

    \item [C)] PnR. After PnR, the final $\mathcal{R}^{i}_{hash}$, $\mathcal{R}^i_{loc}$, $\mathcal{R}^{i}_{rand}$ and $\mathcal{R}^{i}_{repl}$ are obtained. $\mathcal{R}^i_{loc}$ are kept local, $\mathcal{R}^{i}_{rand}$ (\resp $\mathcal{S}^{i}_{rand}$) are randomly redistributed. Therefore, the redistribution cost of PnR can be estimated as
\begin{equation}
\begin{split}
    Re_{PnR} & =\sum_{x\in{\mathcal{R}_{loc}}}{|M_{l}(R,x)|} + \sum_{x\in{\mathcal{R}_{rand}}}{|M_{r}(R,x)|} \\
    & + \sum_{x\in{\mathcal{S}_{rand}}}{|M_{r}(S,x)|}
\end{split}
\end{equation}

    \end{itemize}

\btitle{Join.}
Following the popular formula proposed in \cite{dewitt1992practical}, we estimate the cost of the join on the $i$-th node, \ie $Comp^i$, and
\begin{equation}
Comp^i=|\mathcal{R}^i|_{est}+|\mathcal{S}^i|_{est}+|\mathcal{R}^i \bowtie{\mathcal{S}^i}|_{est}
\end{equation}
$|\mathcal{R}^i|_{est}$
and $|\mathcal{S}^i|_{est}$ are the estimated number of skewed tuples of $R$ and $S$ that are redistributed to $node_i$, respectively, the sum of which reflects the cost of building and probing phases at node $i$.  $|\mathcal{R}^i \bowtie{\mathcal{S}^i}|_{est}$ is the estimated number of the results produced by the join of the skewed tuples, which is the estimation of the materialization cost of join results. $Comp^i$ differs among sub-operators as different redistribution plans are adopted respectively. Notably, no matter which specific sub-operator is adopted, $|\mathcal{R}^i|_{est}$ and $|\mathcal{S}^i|_{est}$ are predictable via statistics of skew values. 

In the follows, we shall study the distribution of tuples with value $x$ after redistribution. 

For $M_h$, tuples with value $x$ are hash redistributed to the $i$-th, $hash(x)\bmod{N}=i$. The computation cost of the value $x$ on the $i$-th node is $|M_h(x)| = Q(R,x) + Q(S,x) + Q(R,x)Q(S,x)$. 

For $M_l$, tuples with value $x$ from $R^i$ are kept local, the tuples with value $x$ from $S^i$ is broadcast. Therefore, the the computation cost of the value $x$ from $R$ on the $i$-th node is $|M^i_l(R, x)| = Q^i(R,x) + Q(S,x) + Q^i(R,x)Q(S,x)$. Similarly, $|M^i_l(S, x)| = Q(R,x) + Q^i(S,x) + Q(R,x)Q^i(S,x)$. 

For $M_r$, tuples with value $x$ from $R^i$ are randomly assigned to each node with the uniform probability of $\frac{1}{N}$. Meanwhile, tuples with value $x$ from $S^i$ are broadcast. Obviously, according to this method, the cost of each node with respect to the value $x$ is the same. Therefore, the the computation cost of the value $x$ from $R$ on the every node is $|M_r(R, x)| = \frac{Q(R,x)}{N} + Q(S,x) + \frac{Q(R,x)}{N}Q(S,x)$. Similarly, $|M_r(S, x)| = \frac{Q(S,x)}{N} + Q(R,x) + \frac{Q(S,x)}{N}Q(R,x)$. 

Next, we conclude and show formally how the computation cost of the three sub-operators are modeled:
\begin{itemize}
\item [A)] GraHJ.
In GraHJ, We divide all skewed (including both partial and complete skewed) values in $R$ and $S$ into $N$ sets, and $R_i \bigcup S_i$ denote the $i$-th set. The values $x \in R_i \bigcup S_i$ follow $hash(x)\bmod{N}=i$. Therefore, the computation cost of the $i$-th node is $\sum_{x\in{R_i \bigcup S_i}}{|M_h(x)|}$. In all, the computation cost of GraHJ is estimated as
\begin{equation}
Comp_{GraHJ}=\max\limits_{i}{\sum_{x\in{R_i \bigcup S_i}}{|M_h(x)|}}
\end{equation}

\item [B)] PRPD. Here, $\rho_a^p(R)$ and $\rho_a^p(S)$ are always kept local, so the computation cost of PRPD is 
\begin{equation}
\begin{split}
        Comp_{PRPD} = \max\limits_{i}\left(\sum_{x\in{\rho_a^p(R)}}{|M_l(R^i, x)|} +\sum_{x\in{\rho_a^p(S)}}{|M_l(S^i, x)|}\right)
\end{split}
\end{equation} 

\item [C)] PnR. According to PnR, $\mathcal{R}^i_{loc}$ are kept local, $\mathcal{R}^{i}_{rand}$ (\resp $\mathcal{S}^{i}_{rand}$) are randomly redistributed. Therefore, the join cost of PnR can be estimated as
\begin{equation}
\begin{split}
Comp_{PnR} & =\max\limits_{i}{\sum_{x\in{\mathcal{R}_{loc}}}{|M^i_l(R, x)|}} \\
& + \sum_{x\in{\mathcal{R}_{rand}}}{|M_r(R, x)|} + \sum_{x\in{\mathcal{S}_{rand}}}{|M_r(S, x)|}
\end{split}
\end{equation}
\end{itemize}

\btitle{Merge.} 
According to~\autoref{ssec42}, the merge cost can be modeled by the number of tuples transmitted to the gateway node over the network. For the first type of execution plan, where local join results on each node need to be gathered to the gateway node, the merge cost of the $i$-th node is $Summ^i=|\mathcal{R}^i \bowtie{\mathcal{S}^i}|_{est}$. Assuming that the gateway is the $j$-th node, the total merge cost can be modeled as $Summ=\sum^{N}_{i,i\neq{j}}{|\mathcal{R}^i \bowtie{\mathcal{S}^i}|_{est}}$. For the second type of subsequent operator, which does not require the global join results to be materialized, the merge cost $Summ$ is $0$.

The total cost for each sub-operator is the summation of the three phases, namely
\begin{equation}
    C=Re+Comp+Summ
\end{equation} 
The cost-based dispatcher calculates the cost for each sub-operator, \ie GraHJ, PRPD and PnR, and adopts the one with the lowest cost. This cost model can be generally applied for \textsf{Dist-HJ} in major shared-nothing systems, \eg KaiwuDB, as long as the stats of tuples read on each node and the statistics towards skewed values are available. 
We further study the time complexity of the proposed cost-based dispatcher. Consider a skewed value $x$, no matter what kind of redistribution plan is adopted, the cost estimation itself requires a constant time. Therefore, when estimating the cost of a particular sub-operator, its complexity is linear with the number of skewed values, say $k$, then the total time complexity is $O(3k)$. In fact, in our follow-up experiments, the overhead of cost-based dispatcher is trivial and negligible empirically.

\begin{algorithm}
\renewcommand{\algorithmicrequire}{\textbf{Input:}} 
\renewcommand{\algorithmicensure}{\textbf{Output:}}
\caption{sub-operator dispatcher}\label{adaptAlgorithm}
\begin{algorithmic}[1]
\Require Each partition of the build table $S$, $S^i$; Each partition of the probe table $R$, $R^i$; The skewed value set of $R$, $\rho_a^p(R)$; The skewed value set of $S$, $\rho_b^p(S)$;
\Ensure The best sub-operator, $selectedSubOperator$

\State $selectedSubOperator\gets NULL$
\State $minCost\gets{\infty}$
\For{each $SubOperator \in \{PRPD, PnR, GraHJ\}$}
    \State $SubOperator.Init({S^i, R^i, \rho_a^p(R), \rho_a^p(S)})$ 
    \State $Re \gets SubOperator.Redistribution()$  
    \State $Comp \gets SubOperator.Computation()$ 
    \State $Summ \gets 0$
    \If {Need to aggregate results}
        \State $Summ \gets SubOperator.Summarization()$
    \EndIf
    \State $Cost \gets Re + Comp + Summ$
    \If {$Cost < minCost$}
        \State $selectedSubOperator \gets SubOperator$
        \State $minCost \gets Cost$ 
    \EndIf
\EndFor
\State \textbf{return} $selectedSubOperator$

\end{algorithmic}
\end{algorithm}

Algorithm~\ref{adaptAlgorithm} shows the pseudocode of cost-based dispatcher. Lines 3-11 calculate the cost of the three sub-operators, respectively. Lines 12-15 finds the particular sub-operator with the lowest cost.
\begin{figure*}[t]
  \centering
  \includegraphics[width=\linewidth]{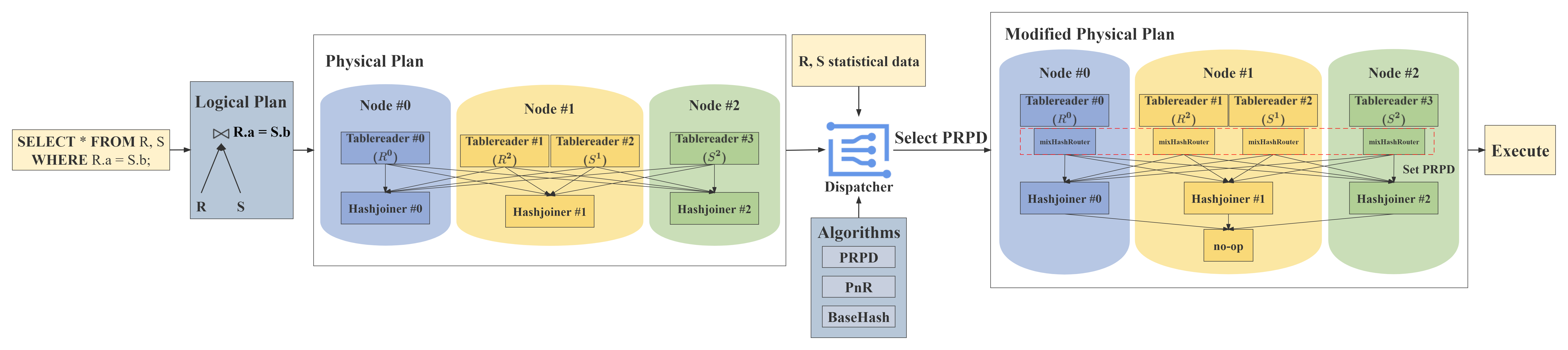}
  \caption{The overall pipeline for executing an SQL involving \textsf{Dist-HJ} (an example showing that the self-adaptive dispatcher finds PRPD)}
  \label{fig:procedure}
\end{figure*} 

\subsection{Implementation of the Dispatcher}\label{ssec5d} 
In the following we describe how PRPD and PnR are implemented and how the sub-operator cost model is applied in KaiwuDB. 

\btitle{Algorithm Implementation.}
    In KaiwuDB, operators are interconnected by streams, which starts from a router connected to the output of the sending operator and ends with a synchronizer connected to the input of the receiving operator. Connecting one \textsf{Tablereader} with all $N$ \textsf{Hashjoiner}s on each node, the router is used to distribute tuples from the \textsf{Tablereader}. By default in KaiwuDB, the router established between \textsf{Tablereader}s and \textsf{Hashjoiner}s only supports \textit{hash redistribution}, \ie $M_h$. In order to add full functional support for PRPD and PnR, a dispatcher is implemented to further enhance the default router, such that it now supports all the four redistribution strategies, \ie \textit{hash}, \textit{broadcast}, \textit{keep local}, \textit{random}. The dispatcher takes the redistribution strategy as a parameter, and distribute tuples according to the selected plan.  
    
\btitle{Cost Model.} 
    The cost is calculated using the statistics on skewed values and the stats of the \textsf{Tablereader}s. The statistics about skewed value can be acquired using table statistics collected by the \textsc{rdbms}. As to the second part, the stats of the \textsf{Tablereader}s can be inferred from the physical plan, which is obtained at the physical planning phase. A physical plan contains the stats of all \textsf{Tablereader}s, which consist of two part:
    \begin{enumerate}
        \item The ID of the nodes each \textsf{Tablereader} is resident on.
        \item The set of records read by each \textsf{Tablereader}.
    \end{enumerate}
    The number of the participating nodes $N$ and the locality of each $R^i$ and $S^i$ can be inferred using (1). The amount of data accessed by \textsf{Tablereader}s are estimated using (2). As discussed in the last section, a range is a chunk of data contiguously partitioned by the primary key of the table. Due to range locality and replica selection, ranges are distributed and accessed via each \textsf{Tablereader}. The amount of data read by each \textsf{Tablereader}, \ie $|R^i|$ and $|S^i|$, can be computed by summing up the volumes of the ranges read by a single \textsf{Tablereader}. So far, all parameters that the cost model requires are obtained.

\autoref{fig:procedure} outlines the pipeline how an SQL involving \textsf{Dist-HJ} in its \textsc{qep} is processed in KaiwuDB under the support of the self-adaptive dispatcher, which eventually adopt the most suitable sub-operator (\eg PRPD) towards the query at runtime.

\section{Evaluation}\label{sec:5}
In this section, we conduct empirical study to test PnR and the self-adaptive cost-based dispatcher. In line with~\cite{2016Flow}, we adopt the same dataset settings to test the skewed \textsf{Dist-HJ}, by generating synthetic datasets that follow the Zipf distribution. Zipf distribution~\cite{zipf2016human} can be used to model the distribution of many variables in real life, such as population income and city size. Zipf distributed datasets are also often used for performance evaluation of database systems, especially the skewness issues~\cite{gray1994quickly}. In the evaluation of \textsf{Dist-HJ} algorithm~\cite{2016Flow}~\cite{kitsuregawa1990bucket}~\cite{1993Using}~\cite{rodiger2014locality} for data skew solutions, Zipf distribution is always employed to compare and justify the performance of the proposed algorithms. Formally, given $n$ elements ranked by their frequency, following a Zipf distribution with skew factor $z$, the most frequent item with rank $1$ shall account for $x = \frac{1}{H_{(n,z)}}$ of all values, where $H_{(n,z)} = \sum_{i=0}^{n}\frac{1}{i^z}$ is the $n$-th generalized
harmonic number. The element with rank $r$ occurs $\frac{x}{r^z}$ times. Besides, we can easily vary the skewness of dataset by changing $z$.

\subsection{Experimental Setup}
In the experiment, we conduct all the tests over a platform with Intel Xeon Gold 6240 $\times 2$, a total of 36 cores, with a memory capacity of 64GB, the persistent storage is SSD, and a network card of Mellanox MT27710 25Gbps. The database version where our self-adaptive cost-based dispatcher is implemented is KaiwuDB ver. 1.0 (ZNBase v2.0), as well as \textsc{crdb} v 20.1.17\footnote{The experimental settings and corresponding results are identical to what we find in KaiwuDB, we select only to show the results in KaiwuDB due to the limit of space and provide our source code for the implementation over \textsc{crdb} at \url{https://github.com/lihuixidian/SkewHJs}}. By default we employ 12 database nodes in the cluster. In terms of database configuration, we expect the tables to be balanced among nodes in the cluster as much as possible, so the range size is set to the minimum value, 65535 bytes. To ensure a fairness comparison among different algorithms for all the tested skew cases, the replica selection of \textsf{Tablereader} is set to random module, where each range is randomly read from 1 out of 3 replicas (nodes) uniformly, and the performance for each algorithm is averaged over 10 runs.

We implement all the three sub-operators as well as our self-adaptive dispatcher, which are listed below, in KaiwuDB following the details shown in \autoref{ssec5d}.
\begin{itemize}
    \item PnR. Partition and Redistribution we propose in \autoref{sec3}.
    \item PRPD. Partial Redistribution Partial Duplication~\cite{2008Handling} with SFR proposed in~\cite{2016Flow}.
    \item GraHJ. Parallel Grace Hash Join proposed in~\cite{DBLP:conf/icde/KitsuregawaTN92}.
    \item Dispatcher. The self-adaptive dispatcher we propose in \autoref{sec:4}.
\end{itemize}
In line with~\cite{2016Flow}, we test the performance for all the baselines using \textit{throughput of tuples}, which refers to the number of tuples processed in the cluster per second. This measure can not only reflect the relative performance of different algorithms, but also the size of the experimental dataset.

\begin{figure*}[t]
    \centering
    \subfigure[3 nodes]{
    \begin{minipage}[t]{0.3\linewidth}
    \centering
    \includegraphics[width=2in]{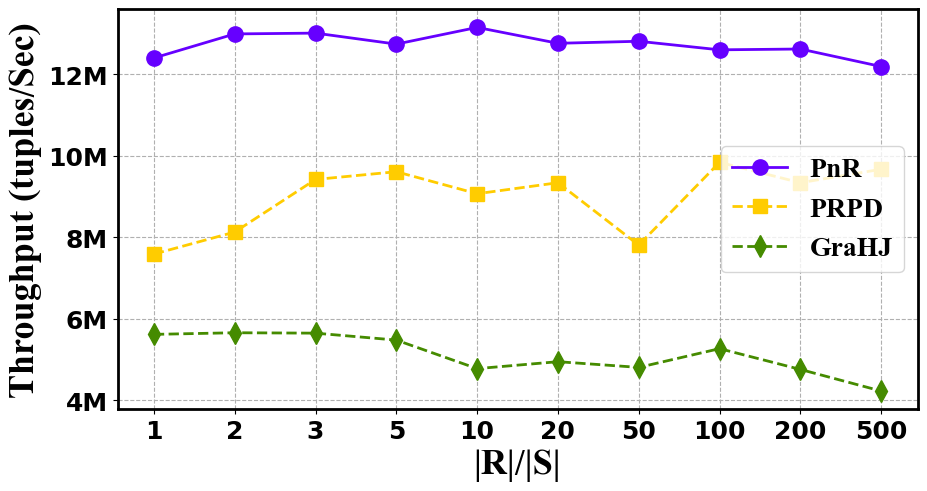}
    \end{minipage}
    }
    \subfigure[6 nodes]{
    \begin{minipage}[t]{0.3\linewidth}
    \centering
    \includegraphics[width=2in]{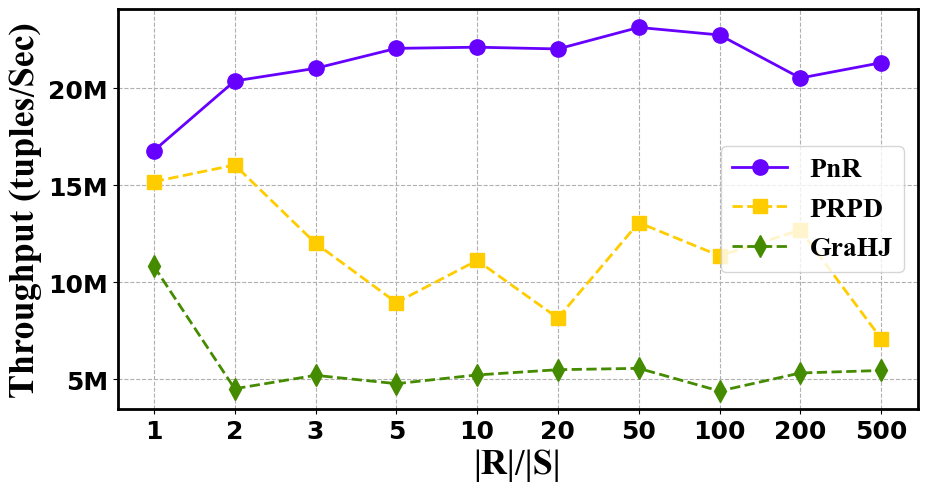}
    \end{minipage}
    }
    \subfigure[12 nodes]{
    \begin{minipage}[t]{0.3\linewidth}
    \centering
    \includegraphics[width=2in]{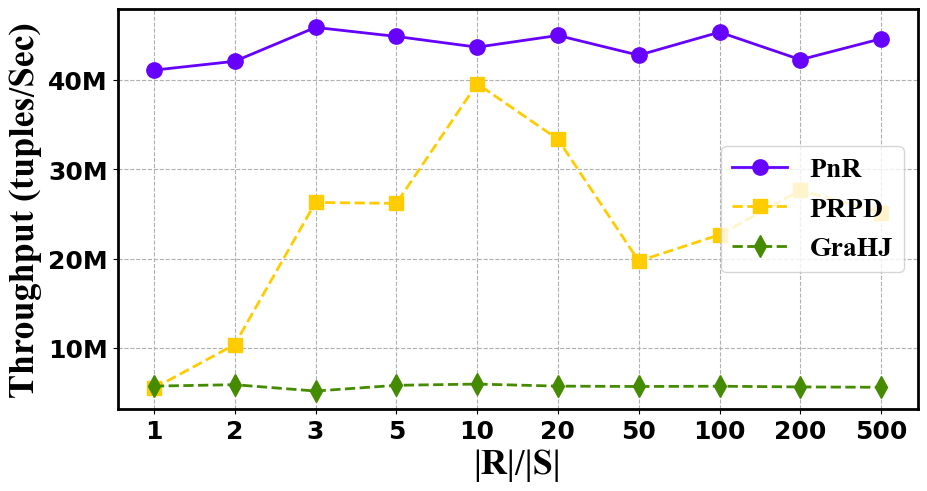}
    \end{minipage}
    }
    \centering
    \vspace{-3ex}\caption{Throughput by increasing relative ratio in the volume of probe and build table}
    \label{fig:RSratio}\vspace{-2ex}
\end{figure*}

\begin{figure*}[t]
    \centering
    \subfigure[3 nodes]{
    \begin{minipage}[t]{0.3\linewidth}
    \centering
    \label{fig:Zipf3}
    \includegraphics[width=2in]{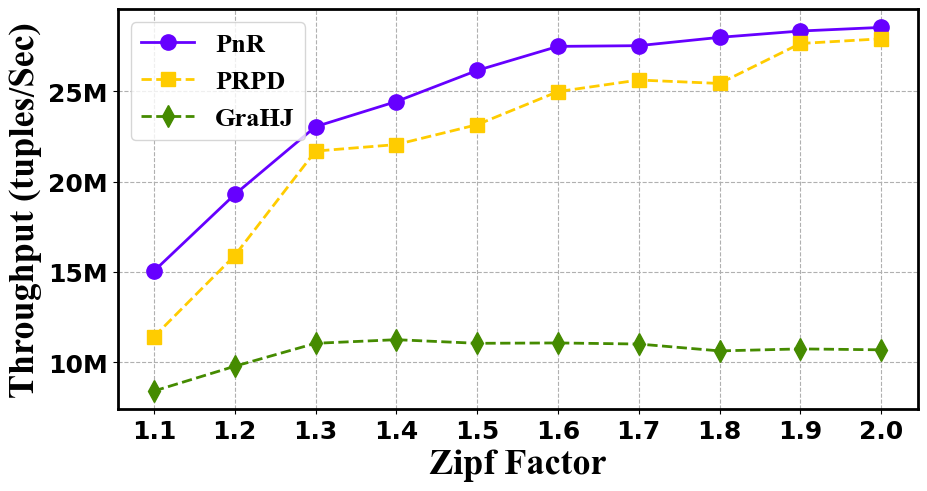}
    \end{minipage}
    }
    \subfigure[6 nodes]{
    \begin{minipage}[t]{0.3\linewidth}
    \centering
    \label{fig:Zipf6}
    \includegraphics[width=2in]{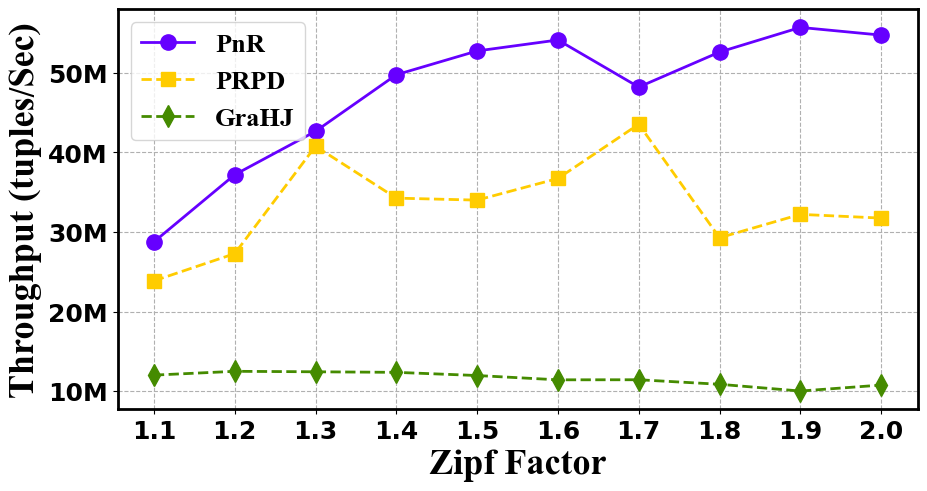}
    \end{minipage}
    }
    \subfigure[12 nodes]{
    \begin{minipage}[t]{0.3\linewidth}
    \centering
    \label{fig:Zipf12}
    \includegraphics[width=2in]{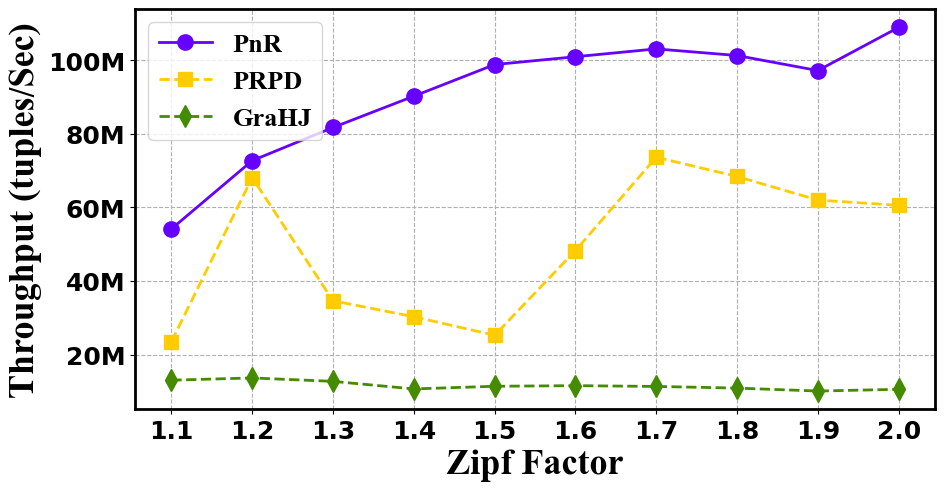}
    \end{minipage}
    }
    \centering
    \vspace{-3ex}\caption{Throughput by increasing Zipf factor $z$ in the probe and build table ($|R|/|S|=293K/10K\;tuples$)}
    \label{fig:Zipf}\vspace{-2ex}
\end{figure*}
\subsection{Performance of PnR}
In order to fully observe the performance of GraHJ, PRPD and PnR, as well as their pros and cons, in the following we conduct experiments by varying the size of joined tables, the degree of skewness, or the number of nodes in the cluster, respectively. 

\btitle{Varying the size of probe/build table.} In the first scenario, the number of nodes is set to fixed values and the parameter of Zipf for both tables are set to $z=1.2$. We fix the size of build table, and vary the size of probe one to tell the performance for different algorithms \wrt the relative ratio in the sizes of joined tables. \autoref{fig:RSratio} shows the throughput for three groups of experiments with different number of nodes. From the experimental results, the performance of PnR is the best and the most stable. On the one hand, because the experiment runs under high-speed network, the network overhead of PnR is limited and negligible. On the other hand, 
randomly redistributed of the PnR ensures stability. The performance of PRPD is unstable, which is caused by the imbalance of data before redistribution during each time of execution.

\btitle{Varying the degree of skewness.} \autoref{fig:Zipf} compares the three algorithms by increasing the Zipf factor $z$. The degree of skewness for the dataset increases as the Zipf factor. When the data becomes more skewed, the performance of PnR will be better. In this part of the experiment, PRPD is still unstable for the same reason as we have discussed in the above. It is worth mentioning that the performance of PRPD on 3 nodes is significantly more stable than that on 6 nodes and 12 nodes. To unveil the secret behind, we investigate the data distribution and find that the smaller the number of nodes in the cluster, the more balanced distribution for the skewed data. At the same time, the empirical results are also consistent with our discussion in~\autoref{sec3} that the theoretical upper limit for the performance of PRPD is close to PnR, such as $z=1.3$ in \autoref{fig:Zipf6} and $z=1.2$ in \autoref{fig:Zipf12}.

\begin{figure}[t]
  \centering
  \includegraphics[width=\linewidth]{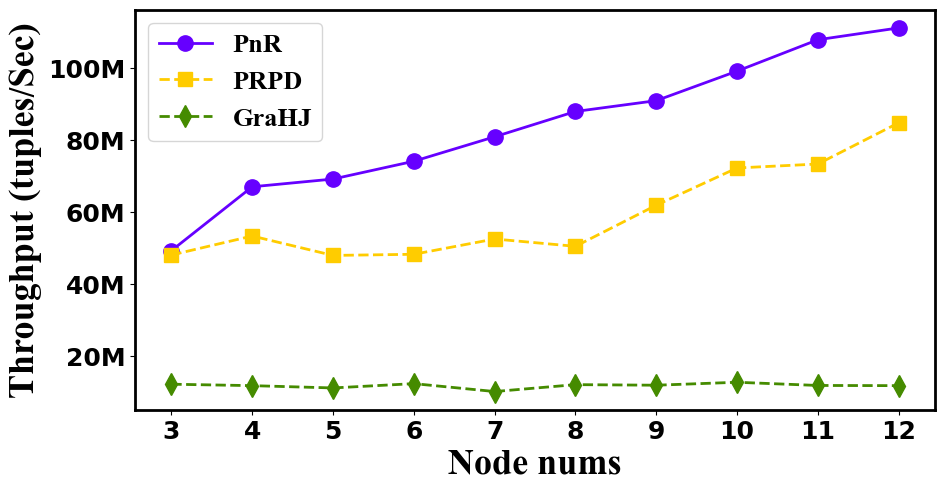}
  \caption{Throughput by increasing the number of nodes within KaiwuDB cluster ($z=1.5,\;|R|/|S|=391K/19K\;tuples$)}
  \label{fig:N}\vspace{-1ex}
\end{figure}

\btitle{Varying the size of the cluster.}
\autoref{fig:N} shows the scalability of the three algorithms when the number of nodes in the cluster increases. Since all skewed tuples with the same value are assigned to the same node, GraHJ gains the least speedup from the increasing number of nodes. Compared with GraHJ, the performance of PRPD has been significantly improved, which can speed up by $4\times$ to $8\times$. PnR has the best scalability and is more stable over different settings.

\subsection{Performance of the Cost-Based Self-adaptive Dispatcher}
In this part, we conduct a group of experiments to test the performance of the Cost-Based Self-adaptive Dispatcher we proposed. In particular, according to our study in \autoref{sec:4}, PnR, PRPD and GraHJ wins in different scenarios, it is easy for our cost-based dispatcher to find the best solution except for the boundary between different scenarios, where the performance of a single algorithm will suddenly change. To this end, our test are conducted targeting these challenging tasks, which we identify as two different experimental settings. Instead of the data set with Zipf distribution, in these experiments we use a single skewed value, in order to control the frequency of skew value more precisely such that we can easily observe the performance change of different \textsf{Dist-HJ} sub-operators whenever a value changes from non-skewed to skewed.

\begin{figure}[t]
  \centering
  \includegraphics[width=\linewidth]{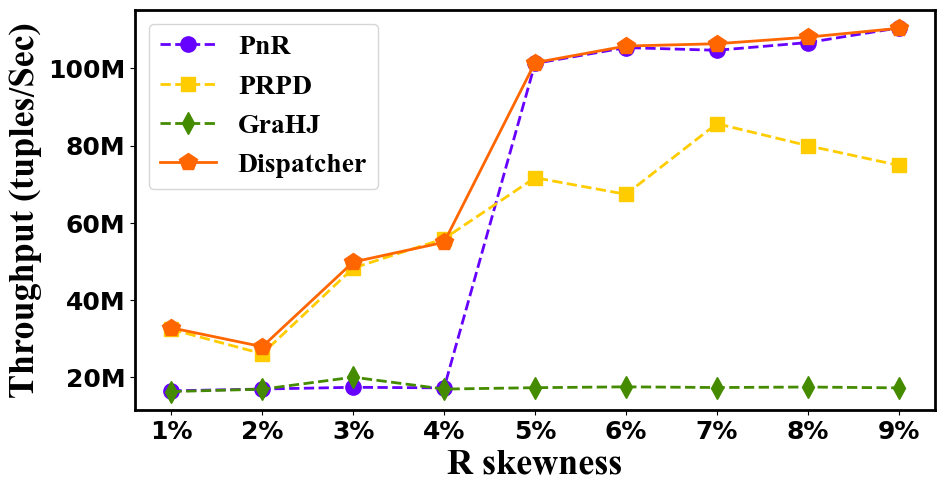}
  \caption{The cost-based dispatcher chooses the optimal sub-operator when the skewness of probe table increases ($12\; nodes,\;|R|/|S|=195K/147K\;tuples$)}
  \label{fig:skewness}\vspace{-1ex}
\end{figure}

\begin{figure}[t]
  \centering
  \includegraphics[width=\linewidth]{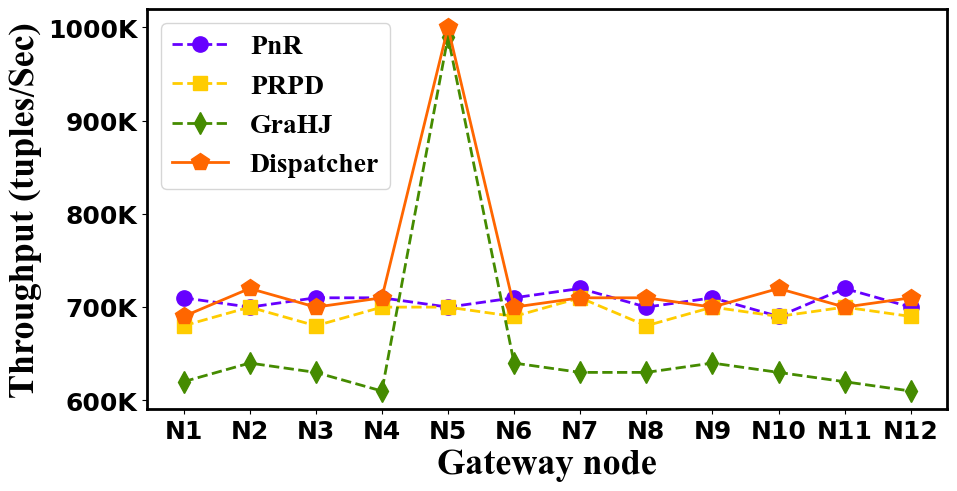}
  \caption{Query execution time for different gateway node ($12\; nodes,\;|R|/|S|=195K/98K\;tuples$)}
  \label{fig:gwn}\vspace{-1ex}
\end{figure}

In the first group of experiments, we set the frequency threshold $p=5\%$, \ie values whose frequencies are above that are viewed as the skewed value. The skewness of the build table is fixed to $50\%$ and the skewness of the probe table increases from $1\%$ to $9\%$. As shown in \autoref{fig:skewness}, the performance of PnR and GraHJ is consistent when the skewness of $R$ is lower than $5\%$. In this case, PnR degenerates to GraHJ because there is no skewed value in $R$ and the PRPD is not affected because it will process the skewed value in $S$. When the the skewness of $R$ is greater than $5\%$, the phenomenon reverses, \ie PnR outperforms PRPD. Nevertheless, no matter PnR beats PRPD or vice versa, Dispatcher always performs the best among all, showing that it correctly finds the best sub-operator at all circumstances.

In the second experimental setting, we evaluate the query time when the same SQL is submitted at different gateway node. In this scenario, the size of the result is $18M$ and need to be summarized by the gateway node. As shown in \autoref{fig:gwn}, when skewed tuples are all redistributed to the gateway node, the query time is $53\%$ faster than the non-gateway node. The cost-based dispatcher is able to identify the optimal sub-operator with the lowest cost, and is superior to the other algorithms.

The cost-based self-adaptive dispatcher performs well in the single skewed workload. In the other workloads like Zipf distribution, similar findings are also observed. In addition, the skew threshold decides how many skewed values can be captured and consequently affects the decision that the dispatcher will make. PnR and PRPD also depend on the skew statistics, thus the self-adaptive dispatcher prefers to choose PnR or PRPD, which is at least as good as GraHJ.

\section{Conclusion and Future Work}\label{sec:6}
Given the fact that skewed data distribution introduces grand challenge towards load balancing in \textsf{Dist-HJ} task, we have conducted plenty of empirical test comparing a series of existing \textsf{Dist-HJ} strategies over different skew settings. We have proposed a novel \textsf{Dist-HJ} strategy, namely PnR, towards a group of representative scenarios where neither GraHJ nor PRPD performs ideally. Our exhaustive study justifies that no single strategy can win in all settings, due to which we proposed a cost-based sub-operator dispatcher and embed it into the cost model and query optimizer such that it can adaptively find the most suitable strategy at runtime. We implement the self-adaptive dispatcher in both \textsc{crdb} and KaiwuDB, empirical results justify that our solution performs the best in all the experimented  settings. Notably, current solution of the self-adaptive dispatcher and corresponding cost model introduced in the end of \autoref{sec:4} relies on the collected statistics for the probe and build table, which can be either the raw relations stored or filtered with an atomic predicate. Intermediate and temporal tables generated during the execution, \ie a temporal results from another precedent join task, cannot be supported by our current solution. To address that, a runtime sampler and estimator should be developed to provide estimation for the input of our self-adaptive dispatcher and the cost model, which plays an important part for our future work in KaiwuDB.
\section*{Acknowledgment}
Thanks for all valuable advice from Dr. Gene Fuh in this project. This work is partially supported by XD-Inspur DB Innovation Lab Grant and National Natural Science Foundation of China (No.61972309, 62272369).

\newpage
\balance
\bibliographystyle{ACM-Reference-Format}
\bibliography{skewedHJ}
\end{document}